\documentclass[journal]{IEEEtran}

\ifCLASSINFOpdf

\else

\fi

\usepackage{mathrsfs}
\usepackage{amsmath}
\usepackage{amsfonts}
\usepackage{amsthm}
\usepackage{amssymb}
\usepackage{graphicx}
\usepackage{subfigure}
\usepackage{indentfirst}
\usepackage{array}
\usepackage{epstopdf}
\usepackage{cite}
\usepackage{mathrsfs}
\usepackage{enumerate}
\usepackage{bm}
\usepackage[linesnumbered,ruled,vlined]{algorithm2e}
\usepackage{setspace}
\usepackage{multirow}
\usepackage{verbatim}
\usepackage{stfloats}
\usepackage{color}
\usepackage[table]{xcolor}
\usepackage{makecell}
\usepackage{cancel}

\SetKwComment{Comment}{//}{}
\SetKwBlock{Begin}{Function}{end}

\begin{document}

\title{Nonlinear Transform Source-Channel Coding for Semantic Communications}

\author{Jincheng~Dai,~\IEEEmembership{Member,~IEEE},
        Sixian~Wang,~\IEEEmembership{Student Member,~IEEE},
        Kailin~Tan,~\IEEEmembership{Student Member,~IEEE},
        Zhongwei~Si,~\IEEEmembership{Member,~IEEE},
        Xiaoqi~Qin,~\IEEEmembership{Member,~IEEE},
        Kai~Niu,~\IEEEmembership{Member,~IEEE},
        and Ping~Zhang,~\IEEEmembership{Fellow,~IEEE}

\thanks{This work was supported in part by the National Natural Science Foundation of China under Grant 92067202, Grant 62001049, Grant 62071058, and Grant 61971062, in part by the Beijing Natural Science Foundation under Grant 4222012. \emph{(Corresponding authors: Jincheng Dai, Ping Zhang)}}

\thanks{J. Dai, S. Wang, K. Tan, Z. Si, and K. Niu are with the Key Laboratory of Universal Wireless Communications, Ministry of Education, Beijing University of Posts and Telecommunications, Beijing 100876, China (e-mail: daijincheng@bupt.edu.cn; sixian@bupt.edu.cn; tankailin@bupt.edu.cn; sizhongwei@bupt.edu.cn; niukai@bupt.edu.cn).}

\thanks{K. Niu is also with Peng Cheng Laboratory, Shenzhen, China.}

\thanks{X. Qin and P. Zhang are with the State Key Laboratory of Networking and Switching Technology, Beijing University of Posts and Telecommunications, Beijing 100876, China (e-mail: xiaoqiqin@bupt.edu.cn; pzhang@bupt.edu.cn).}

\thanks{Our project page can be found at \emph{https://semcomm.github.io/ntscc/}}

\vspace{0em}
}

\maketitle

\begin{abstract}
In this paper, we propose a class of high-efficiency deep joint source-channel coding methods that can closely adapt to the source distribution under the nonlinear transform, it can be collected under the name nonlinear transform source-channel coding (NTSCC). In the considered model, the transmitter first learns a nonlinear analysis transform to map the source data into latent space, then transmits the latent representation to the receiver via deep joint source-channel coding. Our model incorporates the nonlinear transform as a strong prior to effectively extract the source semantic features and provide side information for source-channel coding. Unlike existing conventional deep joint source-channel coding methods, the proposed NTSCC essentially learns both the source latent representation and an entropy model as the prior on the latent representation. Accordingly, novel adaptive rate transmission and hyperprior-aided codec refinement mechanisms are developed to upgrade deep joint source-channel coding. The whole system design is formulated as an optimization problem whose goal is to minimize the end-to-end transmission rate-distortion performance under established perceptual quality metrics. Across test image sources with various resolutions, we find that the proposed NTSCC transmission method generally outperforms both the analog transmission using the standard deep joint source-channel coding and the classical separation-based digital transmission. Notably, the proposed NTSCC method can potentially support future semantic communications due to its content-aware ability and perceptual optimization goal.
\end{abstract}


\IEEEpeerreviewmaketitle

\section{Introduction}\label{section_introduction}

\IEEEPARstart{S}{emantic} communications are recently emerging as a new paradigm driving the in-depth integration of information and communication technology (ICT) advances and artificial intelligence (AI) innovations \cite{zhang2021toward}. This communication paradigm shifting calls for revolutionary theories and methodology innovations. Unlike traditional module-stacking system design philosophy, it is the very time to bridge the two main branches of Shannon information theory \cite{Shannon1948A}, i.e., source and channel parts, together for boosting the end-to-end system capabilities. By this means, the channel transmission process can be aware of the source semantic features.

The paradigm aiming at the integrated design of source and channel codes was named \emph{joint source-channel coding (JSCC)} \cite{verduJSCC}, which is a classical topic in the information theory and coding theory. However, classical JSCC schemes \cite{verduJSCC,guyader2001joint,ramzan2007joint,chen2018joint} were based on the statistical probabilities without considering source semantic aspects. By introducing AI, JSCC is expected to evolve to a modern version. On the one hand, source coding can intelligently extract the most valuable information for human understanding in intelligent human-type communications (IHTC) and decision making in intelligent machine-type communications (IMTC) such that the source coding rate can be efficiently reduced. On the other hand, channel coding can precisely identify the critical part of the source coded sequence to achieve semantics-biased unequal protection.

As one modern version, recent deep learning methods for realizing JSCC have stimulated significant interest in both AI and wireless communication communities \cite{JSCCtext,choi2019neural,DJSCC,DJSCCF,DJSCCL,jankowski2020wireless,ADJSCC}. By using artificial neural networks (ANN), the problem of JSCC can be carried out over analog channels, where transmitted signals are formatted as continuous-valued symbols. From Shannon's perspective \cite{Shannon1948A}, deep JSCC can be viewed as a geometric mapping of a vector in the source space onto a lower-dimensional space embedded in the source space and transmitted over the noisy channel. Similar to the landmark works of G{\"u}nd{\"u}z \cite{DJSCC,DJSCCF,DJSCCL}, we also take the image source as a representative in this paper, but our work is extensible for other source modalities. Standard deep learning JSCC schemes operate on a simple principle: an image, modeled as a vector of pixel intensities $\mathbf x \in {\mathbb R}^m$, is mapped to a vector of continuous-valued channel input symbols $\mathbf s \in {\mathbb R}^k$ via an ANN-based encoding function. We typically have $k < m$, and $R = k/m$ is named \emph{channel bandwidth ratio (CBR)} \cite{DJSCCF} denoting the coding rate. The decoder attempts to recover the source from the channel corrupted sequence ${{\mathbf {\hat s}}}$. This deep JSCC method yields end-to-end transmission performance surpassing classical separation-based JPEG/JPEG2000/BPG source compression combined with ideal channel capacity-achieving code family, especially for sources of small dimensions, e.g., small CIFAR10 image data set \cite{CIFAR10}.



However, one can observe that, in general, as the source dimension increases, e.g., large-scale images, the performance of deep JSCC degrades rapidly, which is even inferior to the classical separation-based coding schemes. In addition, when the CBR $R$ increases, existing deep JSCC schemes cannot provide comparable coding gain as that of classical separated coding schemes, i.e., the slope of performance curve slowing down with the increase of the CBR $R$ or the channel signal-to-noise ratio (SNR). This phenomenon stems from the inherent defect of standard deep JSCC that cannot identify the source distribution for realizing a patch-wise variable-length transmission. For example, when the dimension of embedding vector increases, these embeddings corresponding to simple patches will be saturated rapidly, leading to severe channel bandwidth wastes and inferior coding gain. This saturation phenomenon is more likely to appear on large-scale images that need higher dimensional representation.

Furthermore, current deep JSCC methods do not incorporate any hyperprior as side information, a concept widely used in modern image codecs but unexplored in image transmission using deep JSCC with ANNs.

In this paper, we aim to break the above limits by proposing a new joint source-channel coding architecture that integrates the new concept of nonlinear transform coding \cite{balle2020nonlinear} and deep JSCC, i.e., nonlinear transform source-channel coding (NTSCC). Particularly, as an emerging class of methods, nonlinear transform coding (NTC) over the past few years has become competitive with the best linear transform codecs for image compression, and outperforms them in terms of rate-distortion (RD) performance under well established perceptual quality metrics, e.g., PSNR, MS-SSIM, LPIPS, etc. In contrast to the linear transform coding (LTC) schemes, NTC can more closely adapt to the source distribution, leading to better compression performance. By integrating NTC into deep JSCC, the proposed NTSCC works on the principle: the source vector $\mathbf{x}$ is not directly mapped to the channel input symbols, instead, an alternative (latent) representation of $\mathbf{x}$ is found first, a vector $\mathbf{y}$ in the latent space, and deep JSCC encoding takes place in this latent representation. By introducing an entropy model on the latent space, NTSCC learns a prior as the side information to approximate the distribution of each patch, which is assumed intractable in practice. Accordingly, the following deep JSCC codec can select an appropriate coding scheme that optimizes the transmission RD performance for each embedding $y_i$. As a result, the proposed NTSCC transmission framework can closely adapt to the source distribution and provide superior coding gain. Notably, the proposed NTSCC method can well support future semantic communications due to its content-aware ability and human perceptual optimization goal.

Specifically, the contributions of this paper can be summarized as follows.

\begin{enumerate}[(1)]
  \item \emph{NTSCC Architecture:} We propose a new end-to-end learnable model for high-dimensional source transmission, i.e., NTSCC, that combines the advantages of NTC and deep JSCC. To the best of our knowledge, this is the first work exploiting the nonlinear transform to establish a learnable entropy model for realizing deep JSCC efficiently, where the entropy model on latent code $\mathbf{y}$ implicitly represents the source distribution. In our model, the nonlinear analysis transform condenses the source semantic features as a latent representation $\mathbf{y}$, thus driving the following source-channel coding.

  \item \emph{Adaptive Rate Transmission:} To improve the coding gain of the proposed NTSCC method, we introduce a variable-length transmission mechanism for each embedding vector $y_i$ in the latent code $\mathbf{y}$. To this end, a conditional entropy model $P_{{\bar y}_i | {\mathbf{\bar z}}}$ is performed on each quantized embedding ${\bar y}_i$ to evaluate the entropy of $y_i$. If the learned entropy model indicates the embedding $y_i$ of high entropy, its corresponding deep JSCC shall be assigned a high coding rate, and vice versa. Accordingly, we develop the Transformer ANN architecture and rate attention mechanism to achieve an adaptive rate allocation for deep JSCC, that can finely tune the coding rate for each embedding $y_i$, thus enable source content-aware transmission.

  \item \emph{Hyperprior-aided Codec Refinement:} In the proposed NTSCC method, as the hyperprior on the latent representation, we show that the side information about the entropy model parameters can also be viewed as a prior on the deep JSCC codewords. We exploit this hyperprior to reduce the mismatch between the latent representation marginal distribution for a particular source sample and the marginal for the ensemble of source data the transmission model was designed for. This refinement mechanism only uses a small number of additional bits of information sent from encoder to decoder as signal modifications to achieve much better performance in deep JSCC decoding.

  \item \emph{Performance Validation:} We verify the performance of the proposed NTSCC method across test image sources with various resolutions. Results show that the NTSCC method can achieve much better coding gain and RD performance on various established perceptual metrics such as PSNR, MS-SSIM, and LPIPS \cite{lpips}. Equivalently, achieving the identical end-to-end transmission performance, the proposed NTSCC method can save more than 20\% bandwidth cost, compared to both the emerging analog transmission schemes using the standard deep JSCC and the classical separation-based digital transmission schemes.

\end{enumerate}

The remainder of this paper is organized as follows. In the next section \ref{section_variational}, we first review the variational perspective on deep JSCC and NTC, and propose the variational model for NTSCC. Then, in section \ref{section_architecture}, we propose ANN architectures for realizing NTSCC, as well as key methodologies to guide the optimization of the NTSCC model. Section \ref{section_performance} provides a direct comparison of a number of methods to quantify the performance gain of the proposed method. Finally, section \ref{section_conclusion} concludes this paper.

\emph{Notational Conventions:} Throughout this paper, lowercase letters (e.g., $x$) denote scalars, bold lowercase letters (e.g., $\mathbf{x}$) denote vectors. In some cases, $x_i$ denotes the elements of $\mathbf{x}$, which may also represent a subvector of $\mathbf{x}$ as described in the context. Bold uppercase letters (e.g., $\mathbf{X}$) denote matrices, and $\mathbf{I}_m$ denotes an $m$-dimensional identity matrix. $\ln (\cdot)$ denotes the natural logarithm, and $\log (\cdot)$ denotes the logarithm to base $2$. $p_x$ denotes a probability density function (pdf) with respect to the continuous-valued random variable $x$, and $P_{\bar x}$ denotes a probability mass function (pmf) with respect to the discrete-valued random variable $\bar x$. In addition, $\mathbb{E} [\cdot]$ denotes the statistical expectation operation, and $\mathbb{R}$ denotes the real number set. Finally, $\mathcal{N}(x|\mu, \sigma^2) \triangleq (2\pi \sigma^2)^{-1/2} \exp(-(x - \mu)^2/(2\sigma^2))$ denotes a Gaussian function, and $\mathcal{U}(a-u,a+u)$ stands for a uniform distribution centered on $a$ with the range from $a-u$ to $a+u$.

\section{System Model and Variational Analysis}\label{section_variational}

Consider the following lossy transmission scenario. Alice is drawing a $m$-dimensional vector $\mathbf{x}$ from the source, whose probability is given as $p_{\mathbf{x}}( \mathbf{x} )$. Alice concerns how to map $\mathbf{x}$ to a $k$-dimensional vector $\mathbf{s}$, where $k$ is referred to as the \emph{channel bandwidth cost}. Then, Alice transmits $\mathbf{s}$ to Bob via a realistic communication channel, who uses the received information $\mathbf{\hat s}$ to reconstruct an approximation to $\mathbf{x}$.

\subsection{System Model of Deep JSCC}

\begin{figure}[t]
	\setlength{\abovecaptionskip}{0.cm}
	\setlength{\belowcaptionskip}{-0.cm}
	\centering{\includegraphics[scale=0.37]{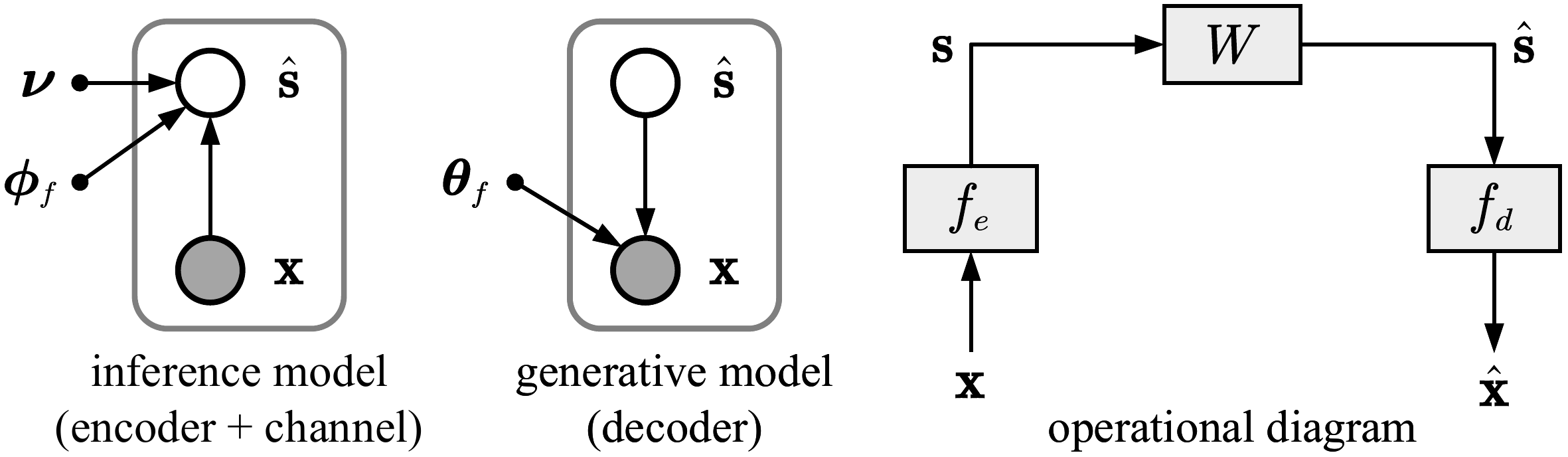}}
	\caption{Left: representation of a deep JSCC encoder combined with the communication channel as a inference model, and corresponding decoder as a generative model. Nodes denote random variables or parameters, and arrows show conditional dependence between them. Right: diagram showing the operational structure of the deep JSCC transmission model. Arrows indicate the data flow, and boxes represent coding functions of data and channel.}\label{Fig1}
	\vspace{0em}
\end{figure}

As stated in the introduction part, in deep JSCC \cite{DJSCC}, the source vector $\mathbf x \in {\mathbb R}^m$, is mapped to a vector of continuous-valued channel input symbols $\mathbf s \in {\mathbb R}^k$ via an ANN-based encoding function ${\mathbf s} = f_{e}( {\mathbf x}; {\bm \phi}_f )$, where the encoder was usually parameterized as a convolutional neural network (CNN) with parameters ${\bm \phi}_f$.  Then, the analog sequence $\mathbf{s}$ is directly sent over the communication channel. The channel introduces random corruptions to the transmitted symbols, denoted as a function $W( \cdot ; \bm{\nu} )$, the channel parameters are encapsulated in $\bm{\nu}$. Accordingly, the received sequence is ${\mathbf{\hat s}} = W( \mathbf{s} ; \bm{\nu} )$, whose transition probability is ${{p_{{\mathbf{\hat s}}| {\mathbf{s}} }}( {{\mathbf{\hat s}}| \mathbf{s} } )}$. In this paper, we consider the most widely used AWGN channel model such that the transfer function is ${\mathbf{\hat s}} = W( \mathbf{s} ; \sigma_n ) = \mathbf{s} + \mathbf{n}$ where each component of the noise vector $\mathbf{n}$ is independently sampled from a Gaussian distribution, i.e., $\mathbf{n} \sim \mathcal{N}(0, {\sigma_n^2}{\mathbf{I}}_k)$, where ${\sigma_n^2}$ is the average noise power. Other channel models can also be similarly incorporated by changing the channel transition function. The receiver also comprises a parametric function ${{\mathbf {\hat x}}} = f_{d}( {{\mathbf {\hat s}}}; {\bm \theta}_f )$ to recover the corrupted signal ${\mathbf{\hat s}}$ as ${{\mathbf {\hat x}}}$, where $f_d$ can also be a format of CNN \cite{DJSCC}. The whole operation is shown in the right panel of Fig. \ref{Fig1}. The encoder and decoder functions are jointly learned to minimize the average
\begin{equation}\label{eq_jscc_target}
  \left( {{{\bm{\phi}}_f^*},{{\bm{\theta}}_f^*}} \right) = \arg \mathop {\min }\limits_{{\bm{\phi}}_f ,{\bm{\theta}}_f } {{\mathbb{E}}_{\mathbf{x}\sim{p_{\mathbf{x}}}}}{{\mathbb{E}}_{{{\mathbf{\hat{x}}} \sim p_{{\mathbf{\hat{x}}} | {\mathbf{x}} }}}}\left[ {d\left( {{\mathbf{x}},{\mathbf{\hat{x}}}} \right)} \right],
\end{equation}
where $d(\cdot; \cdot)$ denotes the distortion loss function.

From the variational perspective, deep JSCC can be viewed as a variational autoencoder (VAE) \cite{kingma2013auto,saidutta2021joint}. As shown in the left panel of Fig. \ref{Fig1}, the noisy sequence $\mathbf{\hat s}$ can be viewed as a sample of latent variables in the generative model. The deep JSCC decoder acts as the generative model (``generating'' the reconstructed source from the latent representation) that transforms a latent variable with some predicted latent distribution into a data distribution that is unknown.

\subsection{Variational Modeling of NTC}

As indicated in the introduction section, when the dimension of representation $k$ becomes large, standard deep JSCC cannot provide sufficient coding gain. As an emerging lossy compression model \cite{balle2016,balle2018,HIFIC,DVC,DCVC}, NTC imitates the classical source coding procedure, that operates on a simple principle: a source vector $\mathbf{x}$ is not mapped to codeword vector directly, rather, an alternative latent representation of $\mathbf{x}$ is found first, a vector in some other space $\mathbf{y}$, quantization then takes place in this latent representation, yielding discrete-valued vector $\mathbf{\bar y}$. In standard NTC, due to the quantization step, the resulting $\mathbf{\bar y}$ can be compressed using entropy coding methods, e.g., arithmetic coding \cite{witten1987arithmetic}, to create bits streams. Since entropy coding relies on a prior probability of $\mathbf{\bar y}$, the \emph{entropy model} $P_{\mathbf{\bar y}}$ is established in NTC to provide side information.

The encoder in NTC transforms the source image vector $\mathbf{x}$ using a parametric nonlinear analysis transform $g_a(\mathbf{x}; \bm{\phi}_g)$ into a latent representation $\mathbf{y}$, which is quantized as $\mathbf{\bar y}$. The latent representation $\mathbf{y}$ preserves the source semantic features while its dimension $n$ is usually much smaller than the source dimension $m$. The decoder performs the inverse operation that recovers $\mathbf{\bar y}$ from the compressed signal first, a parametric nonlinear synthesis transform $g_s(\mathbf{\bar{y}};\bm{\theta}_g)$ is then performed on $\mathbf{\bar y}$ to recover the source image $\mathbf{\hat{x}}$. Here, an ideal error-free transmission is assumed such that the decoder can losslessly recover $\mathbf{\bar y}$ from entropy decoding. In NTC, $g_a$ and $g_s$ are usually parameterized as ANNs of nonlinear property, rather than the conventional linear transforms in LTC, $\bm{\phi}_g$ and $\bm{\theta}_g$ encapsulate their neural network parameters.

Since the nonlinear transform is not strictly invertible, also the quantization step introduces error, the optimization of NTC is attributed to a compression RD problem \cite{berger2003rate}. Assuming an efficient entropy coding is used, rate $R$ as the expected length of the compressed sequence equals to the entropy of $\mathbf{\bar y}$, which is determined by the entropy model $P_{\mathbf{\bar y}}$ as
\begin{equation}\label{eq_ntc_rate_original}
  R = \mathbb{E}_{\mathbf{x}\sim p_{\mathbf{x}}} \left[ -\log{P_{\mathbf{\bar y}}(Q(g_a(\mathbf{x};\bm{\phi}_g)))} \right],
\end{equation}
where $Q$ denotes the quantization function. In the context of this paper, without loss generality, $Q$ employs the uniform scalar quantization $\lfloor \cdot \rceil$ (rounding to integers). Distortion is the expected divergence between $\mathbf{\hat x}$ and $\mathbf{x}$. Clearly, a higher rate allows for a lower distortion, and vice versa. In order to use the gradient descent methods to optimize the NTC model, Ball\'{e} et al. have proposed a relaxed method for addressing the zero gradient problem caused by quantization \cite{balle2016}. It uses a proxy ``uniformly-noised'' representation $\mathbf{\tilde y}$ to replace the quantized representation $\mathbf{\bar y} = \lfloor \mathbf{y} \rceil$ during the model training.

\begin{figure}[t]
	\setlength{\abovecaptionskip}{0.cm}
	\setlength{\belowcaptionskip}{-0.cm}
	\centering{\includegraphics[scale=0.36]{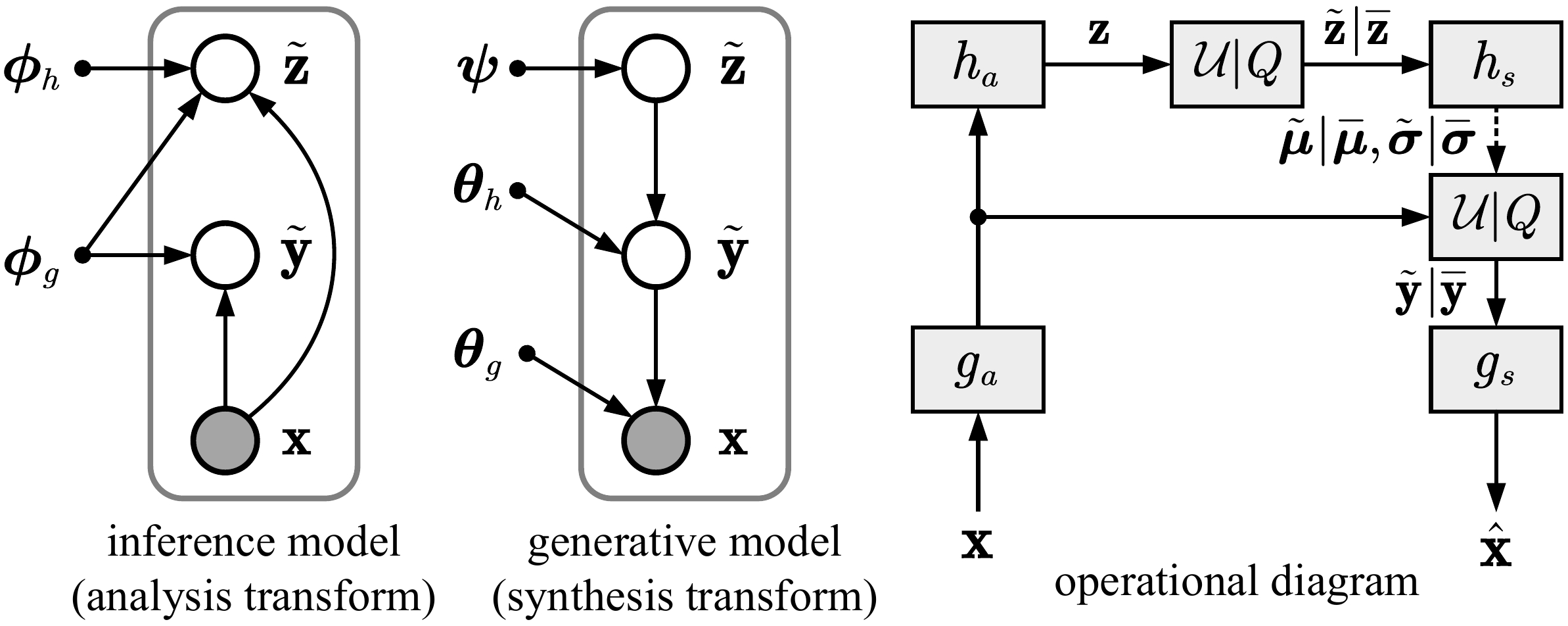}}
	\caption{Left: representation of an NTC encoder as a inference model, and corresponding NTC decoder as a generative model. Nodes denote random variables or parameters, and arrows show conditional dependence between them. Right: diagram showing the operational structure of the NTC compression model. Arrows indicate the data flow, and boxes represent the data transform. Boxes labeled $\mathcal{U} | Q$ denote either uniform noise addition during the model training, or quantization during the model testing.}\label{Fig2}
	\vspace{0em}
\end{figure}

The optimizing problem of NTC can also be formulated as a VAE model as shown in Fig. \ref{Fig2}, a probabilistic generative model stands for the synthesis transform, and an approximate inference model corresponds to the analysis transform. Like that in deep JSCC, the goal of inference model is also creating a parametric variational density $q_{\mathbf{\tilde y} | \mathbf{x}}$ to approximate the true posterior $p_{\mathbf{\tilde y} | \mathbf{x}}$, which is assumed intractable, by minimizing their Kullback-Leibler (KL) divergence over the source distribution $p_{\mathbf{x}}$, i.e.,
\begin{equation}\label{eq_ntc_vae_target}
\begin{aligned}
  & \min\limits_{{\bm \phi}_g, {\bm \theta}_g} \mathbb{E}_{\mathbf{x}\sim p_{\mathbf{x}}} D_{\rm{KL}} \left[q_{\mathbf{\tilde y} | \mathbf{x}} \| p_{\mathbf{\tilde y} | \mathbf{x}} \right] = \min\limits_{{\bm \phi}_g, {\bm \theta}_g} \mathbb{E}_{\mathbf{x}\sim p_{\mathbf{x}}} \mathbb{E}_{\mathbf{\tilde y}\sim q_{\mathbf{\tilde y} | \mathbf{x}}} \\
  & \Big[ \cancelto{\rm{const}_2}{\log{q_{\mathbf{\tilde {y}} | \mathbf{x}}( \mathbf{\tilde{y}} | \mathbf{x})}}  \underbrace{- \log{p_{\mathbf{\tilde{y}}}(\mathbf{\tilde{y}})}}_{\text{rate}}  \underbrace{- \log p_{\mathbf{x} | \mathbf{\tilde y}} ( \mathbf{x} | \mathbf{\tilde{y}} )}_{\text{weighted distortion}} \Big] + \rm{const}_1.
\end{aligned}
\end{equation}
The minimization of KL divergence is equivalent to optimizing the NTC model for \emph{compression RD} performance. As shown in \cite{balle2016}, the first term in \eqref{eq_ntc_vae_target} is computing the transition probability from the source $\mathbf{x}$ to the proxy latent representation $\mathbf{\tilde y}$ as
\begin{equation}\label{eq_ntc_proxy_trans_prob}
  q_{\mathbf{\tilde {y}} | \mathbf{x}}( \mathbf{\tilde{y}} | \mathbf{x}) = \prod_i \mathcal{U}({\tilde y}_i | y_i - \frac{1}{2}, y_i + \frac{1}{2}) \text{~with~} \mathbf{y} = g_a(\mathbf{x}; \bm{\phi}_g),
\end{equation}
where $\mathcal{U}$ denotes a uniform distribution centered on $y_i$. Since the uniform distribution width is constant, the first term is also constant which can be technically dropped. The last term can also be similarly dropped. The third term representing the log likelihood can also be modeled by measuring the squared error between $\mathbf{\tilde x}$ and $\mathbf{x}$, we have $p_{\mathbf{x} | \mathbf{\tilde y}} \left( \mathbf{x} | \mathbf{\tilde{y}} \right) = \mathcal{N}(\mathbf{x} | \mathbf{\tilde{x}},(2\tau_g)^{-1} \mathbf{I}_m )$ with $\mathbf{\tilde x} = g_s(\mathbf{\tilde y}; \bm{\theta}_g)$ where the squared error is weighted by the hyperparameter $\tau_g$. The second term reflects the cross-entropy between the marginal $q_{\mathbf{\tilde y}}(\mathbf{\tilde y}) = \mathbb{E}_{\mathbf{x}\sim{p_{\mathbf{x}}}} [ q_{\mathbf{\tilde y} | \mathbf{x}} (\mathbf{\tilde y} | \mathbf{x} ) ]$ and the prior $p_{\mathbf{\tilde y}} (\mathbf{\tilde y})$. It represents the cost encoding $\mathbf{\tilde y}$ that is constrained by the entropy model $p_{\mathbf{\tilde y}}$. In \cite{balle2016}, Ball\'{e} et al. modeled the prior using a non-parametric fully-factorized density model as
\begin{equation}\label{eq_ntc_entropy_model_y}
  p_{\mathbf{\tilde y} | \bm{\psi}} (\mathbf{\tilde y} | \bm{\psi}) = \prod_i \left( p_{{y}_i | \bm{\psi}^{(i)}} ({y}_i | \bm{\psi}^{(i)}) * \mathcal{U}(-\frac{1}{2},\frac{1}{2}) \right) ({\tilde y}_i),
\end{equation}
where $\bm{\psi}^{(i)}$ encapsulates all the parameters of $p_{{y}_i | \bm{\psi}^{(i)}}$, the convolutional operation ``$*$'' with a standard uniform distribution is used to better match the prior to the marginal. This model is referred to as a \emph{factorized-prior} model \cite{balle2016}.

However, in general cases, there may still exist clear spatial dependencies among the latent representation $\mathbf{\bar y}$, in which case the performance of the factorized-prior model degrades. To tackle this, Ball\'{e} et al. introduced an additional set of latent variables $\mathbf{\tilde z}$ to represent the dependencies of $\mathbf{\bar y}$ in the same way that the original source $\mathbf{x}$ is transformed to the latent representation $\mathbf{\tilde y}$ \cite{balle2018}. Here, each ${\tilde y}_i$ is variationally modeled as a Gaussian with mean ${\mu}_i$ and standard deviation ${\sigma}_i$, where the two parameters are predicted by applying a parametric synthesis transform $h_s$ on $\mathbf{\tilde z}$ as
\begin{equation}\label{eq_ntc_z_to_y}
\begin{aligned}
  p_{\mathbf{\tilde y} | \mathbf{\tilde z}} (\mathbf{\tilde y} | \mathbf{\tilde z}) = & \prod_i \left( \mathcal{N}({\tilde{\mu}}_i,{\tilde{\sigma}}_i^2) * \mathcal{U}(-\frac{1}{2},\frac{1}{2}) \right) ({\tilde y}_i)\\
  ~ & \text{with~} (\bm{\tilde \mu},\bm{\tilde \sigma}) = {h_s}(\mathbf{\tilde z}; \bm{\theta}_h).
\end{aligned}
\end{equation}
The corresponding analysis transform $h_a$ is stacked on top of $\mathbf{\tilde y}$ creating a joint factorized variational posterior as
\begin{equation}\label{eq_ntc_joint_variational_post}
\begin{aligned}
  & q_{\mathbf{\tilde y},\mathbf{\tilde z}|\mathbf{x}}(\mathbf{\tilde y},\mathbf{\tilde z}|\mathbf{x}) = \prod_i \mathcal{U}({\tilde y}_i | y_i - \frac{1}{2}, y_i + \frac{1}{2}) \cdot \prod_j \\
  & \mathcal{U}({\tilde z}_j | z_j - \frac{1}{2}, z_j + \frac{1}{2}) \text{~with~} \mathbf{ y} = g_a(\mathbf{x}; \bm{\phi}_g), \mathbf{ z} = h_a(\mathbf{y}; \bm{\phi}_h).
\end{aligned}
\end{equation}
Since we do not have prior beliefs about the hyperprior $\mathbf{\tilde z}$, it can be modeled as non-parametric fully factorized density like \eqref{eq_ntc_entropy_model_y}, i.e.,
\begin{equation}\label{eq_ntc_entropy_model_z}
  p_{\mathbf{\tilde z}| \bm{\psi}} (\mathbf{\tilde z}| \bm{\psi}) = \prod_j \left( p_{{z}_j | \bm{\psi}^{(j)}} ({z}_j | \bm{\psi}^{(j)}) * \mathcal{U}(-\frac{1}{2},\frac{1}{2}) \right) ({\tilde z}_j),
\end{equation}
where $\bm{\psi}^{(j)}$ encapsulates all the parameters of $p_{{z}_j | \bm{\psi}^{(j)}}$. The optimization goal \eqref{eq_ntc_vae_target} works out to be changed as:
\begin{equation}\label{eq_ntc_vae_target_z}
  \begin{aligned}
  & \min\limits_{{\bm \phi}_g, {\bm \phi}_h, {\bm \theta}_g,{\bm \theta}_h} \mathbb{E}_{\mathbf{x}\sim p_{\mathbf{x}}} D_{\rm{KL}} \left[q_{\mathbf{\tilde y},\mathbf{\tilde z} | \mathbf{x}} \| p_{\mathbf{\tilde y},\mathbf{\tilde z} | \mathbf{x}} \right] = \min\limits_{{\bm \phi}_g, {\bm \phi}_h, {\bm \theta}_g,{\bm \theta}_h} \mathbb{E}_{\mathbf{x}\sim p_{\mathbf{x}}}  \\
  & \mathbb{E}_{\mathbf{\tilde y},\mathbf{\tilde z}\sim q_{\mathbf{\tilde y},\mathbf{\tilde z} | \mathbf{x}}} \Big[ \cancelto{\rm{const}_2}{\log{q_{\mathbf{\tilde {y}},\mathbf{\tilde {z}} | \mathbf{x}}( \mathbf{\tilde{y}},\mathbf{\tilde{z}} | \mathbf{x})}}  \underbrace{- \log{p_{\mathbf{\tilde{y}} | \mathbf{\tilde{z}}}(\mathbf{\tilde{y}}| \mathbf{\tilde{z}})}}_{\text{compression rate}} \underbrace{ - \log{p_{\mathbf{\tilde z}}(\mathbf{\tilde z})}}_{\text{rate of side info.}}   \\
  & \underbrace{- \log p_{\mathbf{x} | \mathbf{\tilde y}} ( \mathbf{x} | \mathbf{\tilde{y}} )}_{\text{weighted distortion}} \Big] + \rm{const}_1,
\end{aligned}
\end{equation}
where the third term can be viewed as the side information widely used in traditional transform coding schemes. The right panel of Fig. \ref{Fig2} depicts the procedure of how the model is used for data compression. Following the variational analysis, the loss function for training the NTC model is
\begin{equation}\label{eq_ntc_loss_function}
\begin{aligned}
  & L = \mathbb{E}_{\mathbf{x}\sim p_{\mathbf{x}}(\mathbf{x})} \bigg[ \lambda( - \log{p_{\mathbf{\tilde y}|\mathbf{\tilde z}}(\mathbf{\tilde y}|\mathbf{\tilde z})} - \log{p_{\mathbf{\tilde z}}(\mathbf{\tilde z})}) + d(\mathbf{x},\mathbf{\hat{x}}_{\text{NTC}})\bigg] \\
  & \text{with~} \mathbf{\tilde y} = g_a(\mathbf{x}; \bm{\phi}_g) + \mathbf{o}, \mathbf{\tilde z} = h_a(\mathbf{y}; \bm{\phi}_h) + \mathbf{o}, \mathbf{\hat x} = g_s(\mathbf{\tilde{y}};\bm{\theta}_g),
\end{aligned}
\end{equation}
where $\mathbf{o}$ denotes uniformly sampling one random quantization offset per latent dimension.

\subsection{Variational Modeling of the Proposed NTSCC}

We integrate both the advantages of NTC and classical deep JSCC, that is collected under the name nonlinear transform source-channel coding (NTSCC). In the transmitter, the analysis transform in NTC is used as a type of ``precoding'' before the encoding of deep JSCC, which extracts the source semantic features as a latent representation. Deep JSCC then operates on this latent space. The bottom panel of Fig. \ref{Fig3} illustrates how the NTSCC model is used for data transmission. The analysis transform module subjects the input source vector $\mathbf{x}$ to $g_a$, yielding the latent representation $\mathbf{y}$ with spatial varying mean values and standard derivations. The latent code $\mathbf{y}$ is then fed into both the analysis transform $h_a$ and the deep JSCC encoder $f_e$. On the one hand, $h_a$ summarizes the distribution of mean values and standard derivations of $\mathbf{y}$ in the hyperprior $\mathbf{z}$, which is then quantized, compressed, and transmitted as side information. The transmitter utilizes the quantized $\mathbf{\bar z}$ to estimate the mean vector $\bm{\bar \mu}$ and the standard derivation vector $\bm{\bar \sigma}$, and use them to determine the bandwidth ratio to transmit the latent representation. The receiver also utilizes $\bm{\bar \mu}$ and $\bm{\bar \sigma}$ to provide side information to correct the probability estimates for recovering $\mathbf{y}$. On the other hand, $f_e$ encodes $\mathbf{y}$ as the channel-input sequence $\mathbf{s}$, and the received sequence is ${\mathbf{\hat s}} = W( \mathbf{s} ; \bm{\nu} )$.

The optimizing problem of NTSCC can also be formulated as a VAE model as shown in Fig. \ref{Fig3}, a probabilistic generative model stands for the deep JSCC decoder and the synthesis transform, an approximate inference model corresponds to the analysis transform and the deep JSCC decoder. Like the above discussion, the goal of the inference model is also creating a parametric variational density $q_{\mathbf{\hat s}, {\mathbf{\tilde z}} | \mathbf{x}}$ to approximate the true posterior $p_{\mathbf{\hat s}, {\mathbf{\tilde z}} | \mathbf{x}}$, which is assumed intractable, by minimizing their KL divergence over the source distribution $p_{\mathbf{x}}$ as \eqref{eq_ntscc_vae_target}. It can be observed that the minimization of the KL divergence is equivalent to jointly optimizing the nonlinear transform model and the deep JSCC model for the end-to-end \emph{transmission RD} performance.
\begin{figure*}[htbp]
\begin{equation}\label{eq_ntscc_vae_target}
  \begin{aligned}
  & \min\limits_{{\bm \phi}_g, {\bm \phi}_h, {\bm \phi}_f, {\bm \theta}_g,{\bm \theta}_h, {\bm \theta}_f} \mathbb{E}_{\mathbf{x}\sim p_{\mathbf{x}}} D_{\rm{KL}} \left[q_{\mathbf{\hat s},\mathbf{\tilde z} | \mathbf{x}} \| p_{\mathbf{\hat s},\mathbf{\tilde z} | \mathbf{x}} \right] \\
  & \overset{(a)}{=} \min\limits_{{\bm \phi}_g, {\bm \phi}_h, {\bm \phi}_f, {\bm \theta}_g,{\bm \theta}_h, {\bm \theta}_f} \mathbb{E}_{\mathbf{x}\sim p_{\mathbf{x}}} \mathbb{E}_{\mathbf{\hat s},\mathbf{\tilde z}\sim q_{\mathbf{\hat s},\mathbf{\tilde z} | \mathbf{x}}} \Big[ \log{q_{\mathbf{\hat s},\mathbf{\tilde z} | \mathbf{x}} ({\mathbf{\hat s},\mathbf{\tilde z} | \mathbf{x}} )} - \log{p_{\mathbf{\hat s},\mathbf{\tilde z}}(\mathbf{\hat s},\mathbf{\tilde z})} - \log{p_{{\mathbf{x}} | \mathbf{\hat s},\mathbf{\tilde z}} ({\mathbf{x}} | \mathbf{\hat s},\mathbf{\tilde z}) } \Big] + \cancelto{\rm{const}_1}{\mathbb{E}_{\mathbf{x}\sim{p_{\mathbf{x}}}}\log{p_{\mathbf{x}}({\mathbf{x}})} } \\
  &  \overset{(b)}{=} \min\limits_{{\bm \phi}_g, {\bm \phi}_h, {\bm \phi}_f, {\bm \theta}_g,{\bm \theta}_h, {\bm \theta}_f} \mathbb{E}_{\mathbf{x}\sim p_{\mathbf{x}}} \mathbb{E}_{\mathbf{\hat s},\mathbf{\tilde z}\sim q_{\mathbf{\hat s},\mathbf{\tilde z} | \mathbf{x}}} \Big[ \log{q_{\mathbf{\hat s},\mathbf{\tilde z} | {\mathbf{x}}} ({\mathbf{\hat s},\mathbf{\tilde z} | {\mathbf{x}}} )} - \log{p_{\mathbf{\tilde z}}(\mathbf{\tilde z})} - \log{p_{\mathbf{\hat s}|\mathbf{\tilde z}}(\mathbf{\hat s}|\mathbf{\tilde z})} - \log{\mathbb{E}_{\mathbf{ y}\sim p_{\mathbf{ y} | \mathbf{\hat s},{\mathbf{\tilde z}}}} \big[ p_{\mathbf{x}| {\mathbf{ y}}}(\mathbf{x}| {\mathbf{ y}}) \big]} \Big] + \rm{const}_1 \\
  & \overset{(c)}{\le } \min\limits_{{\bm \phi}_g, {\bm \phi}_h, {\bm \phi}_f, {\bm \theta}_g,{\bm \theta}_h, {\bm \theta}_f} \mathbb{E}_{\mathbf{x}\sim p_{\mathbf{x}}} \mathbb{E}_{\mathbf{\hat s},\mathbf{\tilde z}\sim q_{\mathbf{\hat s},\mathbf{\tilde z} | \mathbf{x}}} \Big[ \cancelto{\rm{const}_2}{\log{q_{\mathbf{\hat s},\mathbf{\tilde z} | {\mathbf{x}}} ({\mathbf{\hat s},\mathbf{\tilde z} | {\mathbf{x}}} )}} \underbrace{- \log{p_{\mathbf{\tilde z}}(\mathbf{\tilde z})}}_{\text{rate of side info.}} \underbrace{- \log{p_{\mathbf{\hat s}|\mathbf{\tilde z}}(\mathbf{\hat s}|\mathbf{\tilde z})}}_{\text{transmission rate}} \underbrace{- \mathbb{E}_{\mathbf{ y}\sim p_{\mathbf{ y} | \mathbf{\hat s},{\mathbf{\tilde z}}}} \big[ \log p_{\mathbf{x}| {\mathbf{ y}}}(\mathbf{x}| {\mathbf{ y}}) \big]}_{\text{weighted distortion}} \Big] + \rm{const}_1. \\
\end{aligned}
\end{equation}
\hrulefill
\end{figure*}

\begin{figure}[t]
	\setlength{\abovecaptionskip}{0.cm}
	\setlength{\belowcaptionskip}{-0.cm}
	\centering{\includegraphics[scale=0.38]{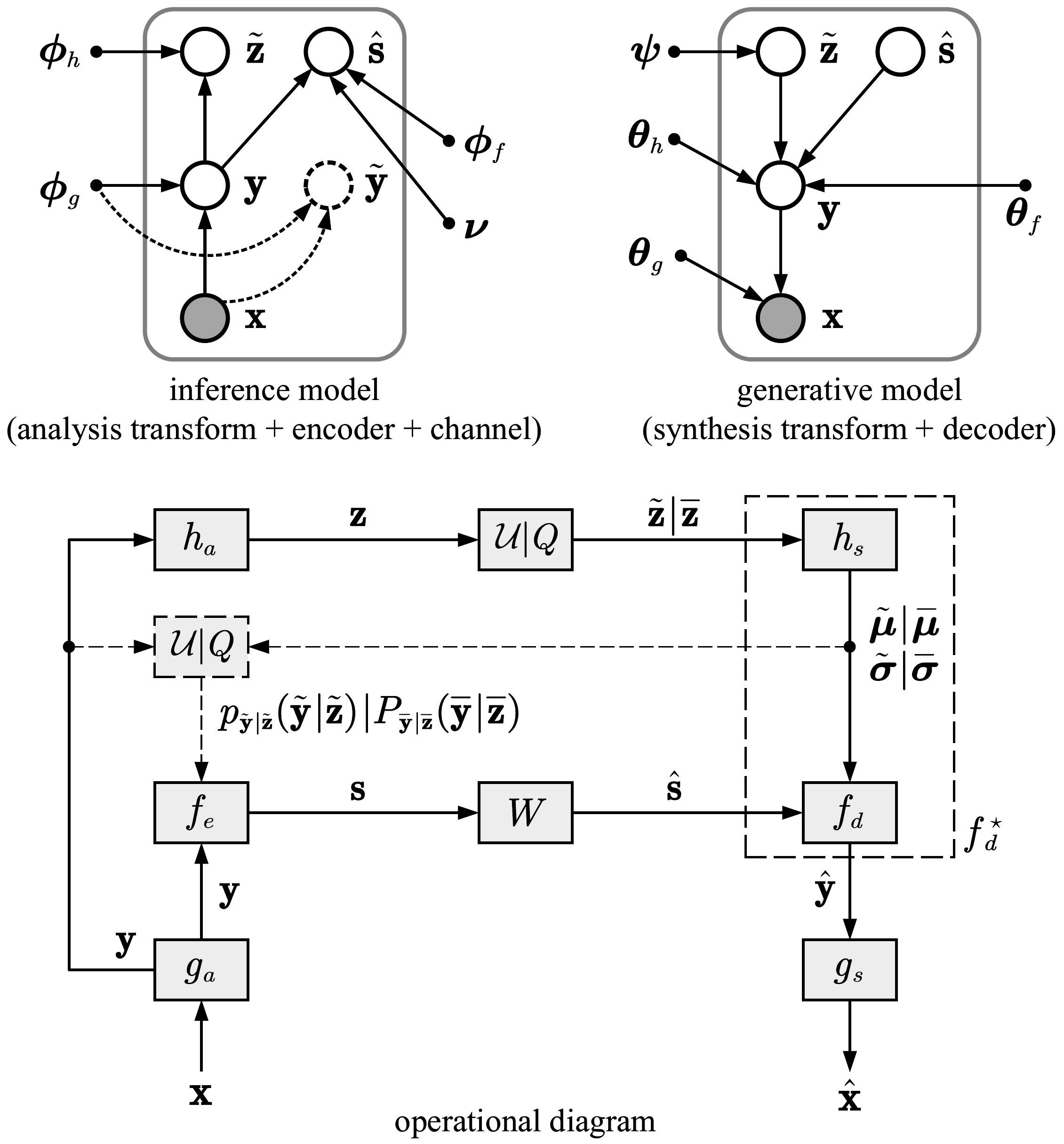}}
	\caption{Top: representation of NTSCC analysis transform encoding combined with the communication channel as a inference model, and corresponding decoding and synthesis transform as a generative model. Nodes denote random variables or parameters, and arrows show conditional dependence between them. Bottom: diagram showing the operational structure of the NTSCC transmission model. Arrows indicate the data flow, and boxes represent the data transform and coding. Boxes labeled $\mathcal{U} | Q$ denote either uniform noise addition during model training, or quantization during model testing. Dashed lines and boxes denotes the calculation of proxy variable.}\label{Fig3}
	\vspace{0em}
\end{figure}

Let's take a close look at each term of the last line in \eqref{eq_ntscc_vae_target}. First, the variational inference model is computing the analysis transform and the deep JSCC encoding of the source vector $\mathbf{x}$, and adding the channel noise $\mathbf{n}$, thus:
\begin{equation}\label{eq_ntscc_term1}
\begin{aligned}
  & q_{\mathbf{\hat s},\mathbf{\tilde z} | \mathbf{x}} ({\mathbf{\hat s},\mathbf{\tilde z} | \mathbf{x}}) = \prod_i \mathcal{N}({\hat s}_i | s_i, \sigma_n^2) \cdot \prod_j \mathcal{U} ({\tilde z}_j | z_j - \frac{1}{2}, z_j + \frac{1}{2})\\
  & \text{with~} \mathbf{y} = g_a(\mathbf{x}; \bm{\phi}_g), \mathbf{s} = f_e(\mathbf{y}; \bm{\phi}_f), \mathbf{z} = h_a(\mathbf{y}; \bm{\phi}_h).
\end{aligned}
\end{equation}
As discussed before, the first term in the KL divergence can be technically dropped from the loss function.

The second term is identical to the cross-entropy between the marginal $q_{\mathbf{\tilde z}}(\mathbf{\tilde z}) = \mathbb{E}_{\mathbf{x}\sim p_{\mathbf{x}}} \mathbb{E}_{\mathbf{\hat s}\sim q_{\mathbf{\hat s} | \mathbf{x}}}  [ q_{\mathbf{\hat s},\mathbf{\tilde z} | \mathbf{x}} ({\mathbf{\hat s},\mathbf{\tilde z} | \mathbf{x}}) ]$ and the prior $p_{\mathbf{\tilde z}} (\mathbf{\tilde z})$. It stands for the cost of encoding the side information in the inference model assuming $p_{\mathbf{\tilde z}}$ as the entropy model. In practice, since we do not have prior beliefs about the hyperprior $\mathbf{\tilde z}$, it can also be modeled as non-parametric fully factorized density as \eqref{eq_ntc_entropy_model_z}.

The third term corresponds to the cross-entropy encoding $\mathbf{\hat s}$ that denotes the transmission rate of source message. As shown in Fig. \ref{Fig3}, we derive $p_{\mathbf{\hat s}|\mathbf{\tilde z}}$ with the help of the intermediate proxy variable $\mathbf{y}$, i.e., the latent presentation of $\mathbf{x}$. Each element $y_i$ is variationally modeled as a Gaussian with mean ${\mu}_i$ and standard deviation ${\sigma}_i$, where the two parameters are predicted by applying a parametric synthesis transform $h_s$ on $\mathbf{\tilde z}$ as
\begin{equation}\label{eq_ntscc_term3_1}
  p_{\mathbf{y}|\mathbf{\tilde z}}(\mathbf{y}|\mathbf{\tilde z}) = \prod_i \mathcal{N}(y_i| {\tilde \mu}_i, {\tilde \sigma}_i^2) \text{~with~} (\bm{\tilde \mu},\bm{\tilde \sigma}) = {h_s}(\mathbf{\tilde z}; \bm{\theta}_h).
\end{equation}
The density $p_{\mathbf{y}|\mathbf{\tilde z}}$ can be transformed to a new density $p_{\mathbf{s}|\mathbf{\tilde z}}$ by using the deep JSCC encoder function as $p_{\mathbf{s}|\mathbf{\tilde z}} = f_e(p_{\mathbf{y}|\mathbf{\tilde z}};\bm{\phi}_f)$ that employs the formula of functional distribution of random variable \cite{jeffreys1998theory}. Under the AWGN channel, the received signal is ${\mathbf{\hat s}} = \mathbf{s} + \mathbf{n}$ with $\mathbf{n} \sim \mathcal{N}(0, {\sigma_n^2}{\mathbf{I}}_k)$. We can thus calculate the density $p_{\mathbf{\hat s}|\mathbf{\tilde z}}$ as
\begin{equation}\label{eq_ntscc_term3_2}
  p_{\mathbf{\hat s}|\mathbf{\tilde z}}(\mathbf{\hat s}|\mathbf{\tilde z}) = \prod_i \left( p_{s_i|\mathbf{\tilde z}}(s_i|\mathbf{\tilde z}) * \mathcal{N}(0,\sigma_n^2) \right) ({\hat s}_i),
\end{equation}
where ``$*$'' denotes the convolutional operation. Since the latent representation $\mathbf{y}$ is directly fed into the deep JSCC encoder without quantization, \eqref{eq_ntscc_term3_2} indeed represents a differential entropy, as opposed to the discrete entropy used for the rate constraint in NTC. Note that $p_{\mathbf{\hat s}|\mathbf{\tilde z}}(\mathbf{\hat s}|\mathbf{\tilde z})$ is originally determined by $p_{\mathbf{y}|\mathbf{\tilde z}}(\mathbf{y}|\mathbf{\tilde z})$, thus, to ensure a stable model training, like that in NTC, we employ a proxy ``uniformly-noised'' representation $\mathbf{\tilde y}$ that is variationally modeled using $\mathbf{\tilde z}$ as \eqref{eq_ntc_z_to_y} to alternate as the transmission rate constraint term in practical implementations of NTSCC, which is marked as dashed lines in the inference model of Fig. \ref{Fig3}. During the model testing, the conditional entropy model $P_{\mathbf{\bar y}|\mathbf{\bar z}}$ is established by taking discrete values from the learned entropy model $p_{\mathbf{\tilde y}|\mathbf{\tilde z}}$, i.e., $P_{\mathbf{\bar y}|\mathbf{\bar z}}({\mathbf{\bar y}|\mathbf{\bar z}}) = p_{\mathbf{\tilde y}|\mathbf{\tilde z}}({\mathbf{\bar y}|\mathbf{\bar z}})$ with $\mathbf{\bar y} = \lfloor \mathbf{y} \rceil$ and $\mathbf{\bar z} = \lfloor \mathbf{z} \rceil$. Therefore, the transmission rate is constrained proportionally to $-\log{P_{\mathbf{\bar y}|\mathbf{\bar z}}(\mathbf{\bar y}|\mathbf{\bar z})}$.

The fourth term represents the log likelihood to recovering $\mathbf{x}$, the output of the synthesis transform $g_s$, which is weighted by the output of the deep JSCC decoder $f_d$. Here, the change of the fourth term from (b) to (c) in \eqref{eq_ntscc_vae_target} follows the law of Jensen inequality, which indicates an upper bound on the KL divergence. The densities $p_{\mathbf{x} | \mathbf{ y}}$ and $p_{\mathbf{ y} | \mathbf{\hat s},{\mathbf{\tilde z}}}$ can be assumed as
\begin{equation}\label{eq_ntscc_term4_1}
  p_{\mathbf{x} | \mathbf{\hat y}} \left( \mathbf{x} | \mathbf{\hat{y}} \right) = \mathcal{N}(\mathbf{x} | \mathbf{\hat x},(2\tau_g)^{-1} \mathbf{I}_m ) \text{~with~} \mathbf{\hat x} = g_s(\mathbf{ y}; \bm{\theta}_g),
\end{equation}
and
\begin{equation}\label{eq_ntscc_term4_2}
\begin{aligned}
  p_{\mathbf{ y} | \mathbf{\hat s},{\mathbf{\tilde z}}} (\mathbf{ y} | \mathbf{\hat s},{\mathbf{\tilde z}}) & \ge p_{\mathbf{y} | \mathbf{\hat s}} (\mathbf{ y} | \mathbf{\hat s}) \cdot p_{\mathbf{y} | \mathbf{\tilde z}} (\mathbf{ y} | \mathbf{\tilde z}) \\
   ~ & = \prod_i \mathcal{N}(y_i | {\check y}_i ,(2\tau_d)^{-1}) \cdot \prod_i \mathcal{N}(y_i | {\tilde \mu}_i ,{\tilde \sigma}_i^2),\\
   ~ & \text{~with~} {\mathbf{\check y}} = f_d(\mathbf{\hat s}; \bm{\theta}_f),(\bm{\tilde \mu},\bm{\tilde \sigma}) = {h_s}(\mathbf{\tilde z}; \bm{\theta}_h),
\end{aligned}
\end{equation}
which measure the squared difference. Different from conventional schemes, \eqref{eq_ntscc_term4_2} indicates that the deep JSCC decoder can use both the received signal $\mathbf{\hat s}$ and the side information $\mathbf{\tilde z}$ to estimate the latent representation $\mathbf{ y}$.

From the above analysis, we also summarize the operations in the receiver, it first recovers the hyperprior $\mathbf{\bar z}$ from the transmitted side information. It then exploits $h_s$ to recover $\bm{\bar \mu}$ and $\bm{\bar \sigma}$, which provide a prior probability helping to recover $\mathbf{\hat y} \doteq \arg \max\limits_{\mathbf{y}} \big[p_{\mathbf{y} | \mathbf{\hat s}} (\mathbf{ y} | \mathbf{\hat s}) \cdot p_{\mathbf{y} | \mathbf{\tilde z}} (\mathbf{ y} | \mathbf{\bar z}) \big]$ that can be computed as \eqref{eq_ntscc_term4_2}. In practice, the decoder for recovering $\mathbf{\hat y}$ can be summarized as a new parametric function $\mathbf{\hat y} = f_d^{\star}(\mathbf{\hat s}, \bm{\bar \mu}, \bm{\bar \sigma} ; \bm{\theta}_{f^{\star}})$ with both $\mathbf{\hat s}$ and $\bm{\bar \mu}, \bm{\bar \sigma}$ as inputs, where $\bm{\theta}_{f^{\star}}$ encapsulates the ANN parameters constituting $f_d^{\star}$. During model training, $\mathbf{\bar z}$ is replaced by the ``uniformly-noised'' proxy $\mathbf{\tilde z}$ to generate $\bm{\tilde \mu}, \bm{\tilde \sigma}$. It then feeds $\mathbf{\hat y}$ into $g_s$ to reconstruct the source $\mathbf{\hat x}$. The whole transceiver operation diagram is illustrated in Fig. \ref{Fig3}.

\section{Architecture and Implementations}\label{section_architecture}

In this section, we discuss details about the NTSCC architecture and key methodologies. Following the aforementioned VAE model, the optimizing of NTSCC can also be attributed as a transmission RD optimization problem, i.e.,
\begin{equation}\label{eq_ntscc_loss_function}
\begin{aligned}
  & L = \\
  & \mathbb{E}_{\mathbf{x}\sim p_{\mathbf{x}}(\mathbf{x})} \bigg[ \lambda( \underbrace{-\eta \log{P_{\mathbf{\bar y}|\mathbf{\bar z}}(\mathbf{\bar y}|\mathbf{\bar z})}}_{k_y} \underbrace{- \frac{\log{P_{\mathbf{\bar z}}(\mathbf{\bar z})}}{C_z}}_{k_z}) + d(\mathbf{x},\mathbf{\hat{x}}_{\text{NTSCC}})\bigg] \\
  & \text{with~} \mathbf{y} = g_a(\mathbf{x}; \bm{\phi}_g), \mathbf{\bar y} = \lfloor \mathbf{y} \rceil, \mathbf{z} = h_a(\mathbf{y}; \bm{\phi}_h), \mathbf{\bar{z}} = \lfloor \mathbf{z} \rceil, \\
  & \mathbf{\hat{s}} = W( f_e(\mathbf{y}; \bm{\phi}_f) ; \bm{\nu} ), (\bm{\bar \mu}, \bm{\bar \sigma}) = h_s(\mathbf{\bar z}; \bm{\theta}_h), \\
  & \mathbf{\hat y} = f_d^{\star}(\mathbf{\hat s}, \bm{\bar \mu}, \bm{\bar \sigma} ; \bm{\theta}_{f^{\star}}), \mathbf{\hat x} = g_s(\mathbf{\hat{y}};\bm{\theta}_g),
\end{aligned}
\end{equation}
where the parameter $\eta$ controls the \emph{scaling relation} between the entropy of latent representation $\mathbf{\bar y}$ and its analog transmission channel bandwidth cost $k_y$, $C_z$ denotes the digital channel capacity to transmit the quantized hyperprior $\mathbf{\bar z}$, thus the digital channel bandwidth cost $k_z$ can be computed to transmit the side information. The Lagrange multiplier $\lambda$ on the total transmission channel bandwidth term determines the trade-off between the transmission rate $k = k_y + k_z$ and the end-to-end distortion $d$. Moreover, from the above variational modeling analysis, we find that both the conditional entropy model $P_{\mathbf{\bar y}|\mathbf{\bar z}}(\mathbf{\bar y}|\mathbf{\bar z})$ and the hyperprior model $P_{\mathbf{\bar z}}(\mathbf{\bar z})$ can be factorized. In order to use the gradient descent methods to optimize the NTSCC model, we relax the quantized variables as ``uniformly-noised'' proxies like that in NTC, therefore, the loss function for model training can be written as
\begin{equation}\label{eq_ntscc_loss_function_factorized}
\begin{aligned}
  & L = \mathbb{E}_{\mathbf{x}\sim p_{\mathbf{x}}(\mathbf{x})} \bigg[ \lambda \Big( \sum_i \underbrace{-\eta \log{p_{{\tilde y}_i|\mathbf{\tilde z}}({\tilde y}_i|\mathbf{\tilde z})}}_{{\tilde k}_{y_i}} + \\
  & \sum_j \underbrace{-\frac{1}{C_z} \log{p_{{\tilde z}_j | \bm{\psi}^{(j)}}({\tilde z}_j | \bm{\psi}^{(j)} )}}_{{\tilde k}_{z_j}}\Big) + d(\mathbf{x},\mathbf{\hat{x}}_{\text{NTSCC}})\bigg] \text{~with~}\\
  & {\tilde y}_i = y_i + o_i, {\tilde z}_j = z_j + o_j, \mathbf{y} = g_a(\mathbf{x}; \bm{\phi}_g), \mathbf{z} = h_a(\mathbf{y}; \bm{\phi}_h),
\end{aligned}
\end{equation}
where offsets $o_i$ and $o_j$ are sampled from uniform distribution $ \mathcal{U}(-\frac{1}{2},\frac{1}{2})$, $p_{{\tilde y}_i|\mathbf{\tilde z}}$ and $p_{{\tilde z}_j | \bm{\psi}^{(j)}}$ are established by taking values from the parametric factorized model in \eqref{eq_ntc_z_to_y} and the non-parametric factorized model in \eqref{eq_ntc_entropy_model_z}, respectively, as
\begin{equation}\label{eq_ntscc_z_to_y_factorized}
\begin{aligned}
  p_{{\tilde y}_i|\mathbf{\tilde z}} ({\tilde y}_i|\mathbf{\tilde z}) = & \left( \mathcal{N}({\tilde{\mu}}_i,{\tilde{\sigma}}_i^2) * \mathcal{U}(-\frac{1}{2},\frac{1}{2}) \right) ({\tilde y}_i)\\
  ~ & \text{with~} (\bm{\tilde \mu},\bm{\tilde \sigma}) = {h_s}(\mathbf{\tilde z}; \bm{\theta}_h),
\end{aligned}
\end{equation}
and
\begin{equation}\label{eq_ntscc_entropy_model_z_factorized}
  p_{{\tilde z}_j| \bm{\psi}^{(j)}} ({\tilde z}_j| \bm{\psi}^{(j)}) = \left( p_{{z}_j | \bm{\psi}^{(j)}} ({z}_j | \bm{\psi}^{(j)}) * \mathcal{U}(-\frac{1}{2},\frac{1}{2}) \right) ({\tilde z}_j).
\end{equation}
Correspondingly, during model testing, the conditional entropy model $P_{{\bar y}_i|\mathbf{\bar z}}$ is established by taking discrete values from the learned model $p_{{\tilde y}_i|\mathbf{\tilde z}}$ as
\begin{equation}
  P_{{\bar y}_i|\mathbf{\bar z}}({\bar y}_i|\mathbf{\bar z}) = p_{{\tilde y}_i|\mathbf{\tilde z}}({\bar y}_i|\mathbf{\bar z}),
\end{equation}
and $P_{{\bar z}_j| \bm{\psi}^{(j)}}$ is similarly computed as
\begin{equation}
  P_{{\bar z}_j| \bm{\psi}^{(j)}} ({\bar z}_j| \bm{\psi}^{(j)}) = p_{{\tilde z}_j| \bm{\psi}^{(j)}} ({\bar z}_j| \bm{\psi}^{(j)}).
\end{equation}

\subsection{The Overall Architecture of NTSCC}

The learned NTSCC model in \eqref{eq_ntscc_loss_function_factorized} indicates that the pmf $P_{\mathbf{\bar y}|\mathbf{\bar z}}(\mathbf{\bar y}|\mathbf{\bar z})$ for predicting the entropy of latent representation is factorized over each dimension without relying on preceding dimensions. Conditioning on the hyperprior vector $\mathbf{\bar z}$ typically requires transmitting the vector as side information. The whole structure corresponds to a learned forward adaptation (FA) of the density model \cite{balle2020nonlinear}. To seek computational parallelism, in this paper, we focus on the FA mode, a better performance backward adaptation (BA) mode with higher processing latency will be discussed in the future \cite{minnen2018}.

\begin{figure}[t]
	\setlength{\abovecaptionskip}{0.cm}
	\setlength{\belowcaptionskip}{-0.cm}
	\centering{\includegraphics[scale=0.41]{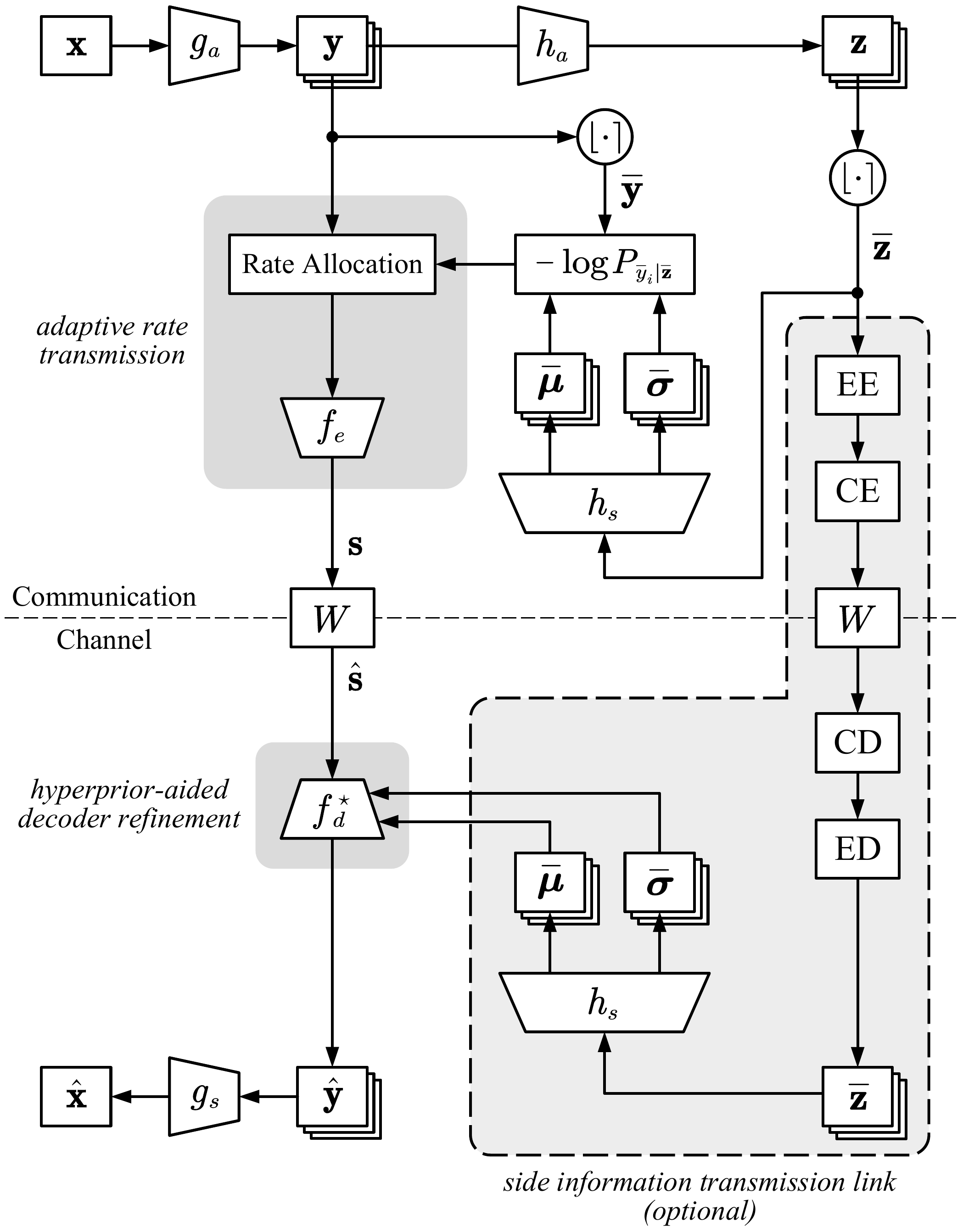}}
	\caption{Illustration of an NTSCC codec architecture using learned FA.}\label{Fig4}
	\vspace{0em}
\end{figure}

Fig. \ref{Fig4} illustrates the overall architecture of NTSCC using learned FA. $\mathbf{x}$ is the source vector at the transmitter, and $\mathbf{\hat x}$ denotes the recovered vector at the receiver. $\mathbf{y}$ is the semantic latent representation tensor that is obtained by performing an ANN-based transform function $g_a$ on $\mathbf{x}$, and $\mathbf{\bar y}$ is the uniformly quantized counterparts of $\mathbf{y}$. Also, $\mathbf{z}$ is the latent representation of $\mathbf{y}$ computed using an ANN-based transform $h_a$, denoting the side information, whose uniformly quantized version is $\mathbf{\bar z}$. While the entropy model of $\mathbf{\bar y}$ on $\mathbf{\bar z}$ is predetermined, the factorized entropy model of $\mathbf{\bar y}$ is assumed to be conditionally independent Gaussian with the mean tensor $\bm{\bar \mu}$ and the standard deviation tensor $\bm{\bar \sigma}$ as \eqref{eq_ntscc_z_to_y_factorized}. The two tensors are obtained by performing the ANN-based function $h_s$ on $\mathbf{\bar z}$. The resulting entropy terms $-\log P_{{\bar y}_i|\mathbf{\bar z}}({\bar y}_i|\mathbf{\bar z})$ are employed for determining the channel bandwidth cost of each dimension $y_i$ so as to achieve adaptive rate transmission. Bob begins with channel decoding (CD) and entropy decoding (ED) to recover the side information $\mathbf{\bar z}$, and then uses it to decode $\mathbf{\hat y}$. Alice should know the entropy model on $\mathbf{\bar z}$ to entropy encode (EE) and channel encode (CE) it, that is modeled as a non-parametric density conditioned on $\bm{\psi}$ as \eqref{eq_ntscc_entropy_model_z_factorized}.

\emph{A Special Note: As shown in Fig. \ref{Fig4}, the side information $\mathbf{\bar z}$ is not necessary for Bob.} As analyzed in previous section, when the hyperprior $\mathbf{\bar z}$ is transmitted through the side information link, to decode $\mathbf{\hat y}$, Bob can jointly use channel received $\mathbf{\hat s}$ and side information $\bm{\bar \mu}, \bm{\bar \sigma}$ as the inputs of decoder function $f_d^{\star}$ for getting better performance. In this case, to ensure reliable transmission of the side information, channel coding (CE and CD) should adopt advanced capacity-approaching codes, such as low-density parity-check (LDPC) codes \cite{LDPC_5G} or polar codes \cite{Polar}. On the other hand, when $\mathbf{\bar z}$ is only used by Alice to allocate channel bandwidth cost of each $y_i$, some performance degradation will be observed in the deep JSCC decoder $f_d$ while the cost for transmitting $\mathbf{\bar z}$ can be omitted. In subsequent sections, unless specified clearly, NTSCC refers to the model where the hyperprior $\mathbf{\bar z}$ is transmitted as the side information to refine the decoder. Detailed performance comparison about whether to transmit the side information $\mathbf{\bar z}$ will be given in the subsequent ablation study.

\begin{algorithm}[t]
\setlength{\abovecaptionskip}{0cm}
\setlength{\belowcaptionskip}{-0cm}
\caption{Training the NTSCC model}\label{alg_training}
\KwIn {Training data $\mathbf{x}$, the Lagrange multiplier $\lambda$ on the rate term, the scaling factor $\eta$ from entropy to channel bandwidth cost, learning rate $\gamma$\;}

\Comment{model pre-training}
Initialize parameters $(\bm{\phi}_g^{(0)},\bm{\phi}_h^{(0)},\bm{\theta}_g^{(0)},\bm{\theta}_h^{(0)})$ by pre-training the corresponding NTC model as that in \cite{balle2020nonlinear}\;

Randomly initialize parameters $(\bm{\phi}_f^{(0)},\bm{\theta}_{f^{\star}}^{(0)})$\;

Encapsulate $\mathbf{w}^{(0)} = (\bm{\phi}_g^{(0)},\bm{\phi}_h^{(0)},\bm{\phi}_f^{(0)},\bm{\theta}_g^{(0)},\bm{\theta}_h^{(0)},\bm{\theta}_{f^{\star}}^{(0)})$\;

\For{$t = 1,2,\cdots,T$}
{
    Sample $\mathbf{x} \sim p_{\mathbf{x}}$\;

    \Comment{add the NTC distortion to improve the training convergence stability}
    Calculate the RD loss function $L(\mathbf{w}^{(t-1)},\lambda,\eta) = d(\mathbf{x},\mathbf{\hat x}_{\text{NTSCC}}) + d(\mathbf{x},\mathbf{\hat x}_{\text{NTC}}) + \lambda(k_y + k_z)$\;

    Calculate the gradients $\nabla_{\mathbf{w}^{(t-1)}} L(\mathbf{w}^{(t-1)},\lambda,\eta)$\;

    Update $\mathbf{w}^{(t)} \leftarrow \mathbf{w}^{(t-1)} - \gamma \nabla_{\mathbf{w}^{(t-1)}} L(\mathbf{w}^{(t-1)},\lambda,\eta)$\;
}
\Return Trained parameters $\mathbf{w}^{(T)} = (\bm{\phi}_g^{(T)},\bm{\phi}_h^{(T)},\bm{\phi}_f^{(T)},\bm{\theta}_g^{(T)},\bm{\theta}_h^{(T)},\bm{\theta}_{f^{\star}}^{(T)})$.\\
\end{algorithm}

\begin{figure*}[t]
	\setlength{\abovecaptionskip}{0.cm}
	\setlength{\belowcaptionskip}{-0.cm}
	\centering{\includegraphics[scale=0.35]{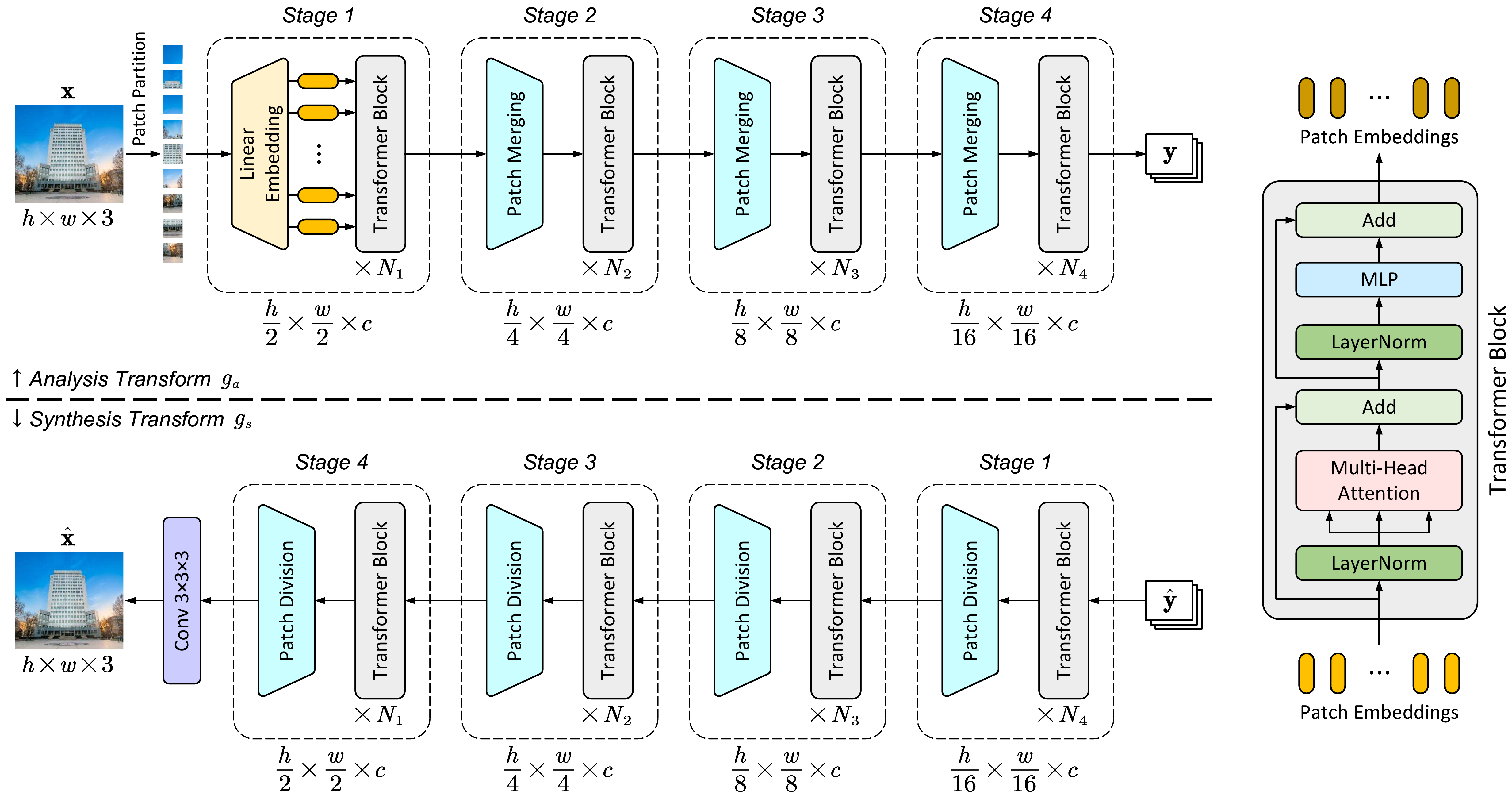}}
	\caption{Network architectures of nonlinear transforms $g_a$ and $g_s$.}\label{Fig5}
	\vspace{0em}
\end{figure*}

Next, we discuss the procedure of NTSCC model training as shown in Algorithm \ref{alg_training}. Here, some tricks should be noted to ensure fast and stable training. First, before the training of NTSCC model, the parameters of $g_a$, $g_s$, $h_a$, $h_s$ should be initialized by pre-training the corresponding NTC model as \eqref{eq_ntc_loss_function}, where no transmitting error is considered such that NTC only executes the data compression task. During the training of NTSCC model, we modify the loss function derived in \eqref{eq_ntscc_loss_function_factorized} by adding the NTC distortion terms, in this way, the convergence of NTSCC model training can be more stable and we can maintain the performance of NTC component for compression task. Moreover, due to the introduction of rate adaptation in NTSCC, the channel bandwidth cost of each embedding $y_i$ is different, thus, multi-head ANN structures shall be used to implement deep JSCC codec $f_e$ and $f_d^{\star}$. In each round of model training, only parts of $\bm{\phi}_f$ and $\bm{\theta}_{f^{\star}}$ will be updated depending on the selected transmitting rates. Details will be given later.

\subsection{Modular Implementation Details}

In this part, we present the implementation details of each module in NTSCC. As aforementioned, the proposed NTSCC model mainly consists of a nonlinear transform step and a rate-adaptive deep JSCC step. The key of NTSCC implementation is designing dynamic and efficient ANN structures that can learn patch-wise representations and use the side information provided by hyperprior to flexibly determine the transmission bandwidth cost of each patch. To this end, the encoder function $f_e$ should incorporate these external parameters, such as the transmission rate of each embedding $y_i$ that is determined by the learned entropy model $-\log P_{{\bar y}_i|\mathbf{\bar z}}({\bar y}_i|\mathbf{\bar z})$ and the predetermined scaling factor $\eta$. We inform JSCC codec ANNs of this global or contextual information using ``\emph{indicators}''. These indicators are generally referred to as a group of learnable parameters, each one is associated with a specific external state. In this work, we employ vision Transformers as the ANN backbone \cite{ViT}, that integrates tokens or positional embedding as indicators.

Next, we describe the implementation details of each module in NTSCC, where the source is assumed as image.

\subsubsection{Nonlinear Transform Modules $g_a$, $g_s$, $h_a$, and $h_s$}

The proposed network architecture of the analysis transform $g_a$ is illustrated in Fig. \ref{Fig5}. An RGB image source $\mathbf{x} \in \mathbb{R}^{h\times w\times3}$ is first split into $l_1=\frac{h}{2} \times \frac{w}{2}$ non-overlapping patches and each patch is of $2 \times 2 \times 3$ dimensions. Each patch embedding can be viewed as a ``token'', we thus have a sequence of tokens $( x_1, \dots, x_{l_1} )$ by putting these tokens in the order from top left to bottom right. After the patch partition, patch tokens are fed into a linear fully-connected (FC) layer for obtaining the embedding with $c$ feature dimensions.

Then, $N_1$ Transformer blocks are applied on these $l_1$ patch embedding tokens \cite{ViT}. As shown in Fig. \ref{Fig5}, a Transformer block is a sequence-to-sequence function that consists of a multi-head self-attention layer and a feed-forward layer with both having skip connection and layer normalization (LayerNorm) operations \cite{liu2021Swin, Layernorm}. The operations in each layer of the Transformer block are described as
\begin{equation}
\mathbf{O}_1 = \mathbf{X} + \rm{MHSA} (\mathbf{X})
\end{equation}
and
\begin{equation}
\mathbf{O}_2 = \mathbf{O}_1 + \rm{MLP}( \rm{LayerNorm}(\mathbf{O}_1)),
\end{equation}
where $\mathbf{O}_1$ and $\mathbf{O}_2$ denote the output from the self-attention layer and the feed-forward layer, respectively. The function of MHSA denotes the multi-head self-attention with learnable relative positional bias \cite{raffel2020exploring}. The function of MLP consists of a hidden layer and an embedding layer with the GELU activation \cite{hendrycks2016gaussian}.

\begin{figure}[t]
	\setlength{\abovecaptionskip}{0.cm}
	\setlength{\belowcaptionskip}{-0.cm}
	\centering{\includegraphics[scale=0.325]{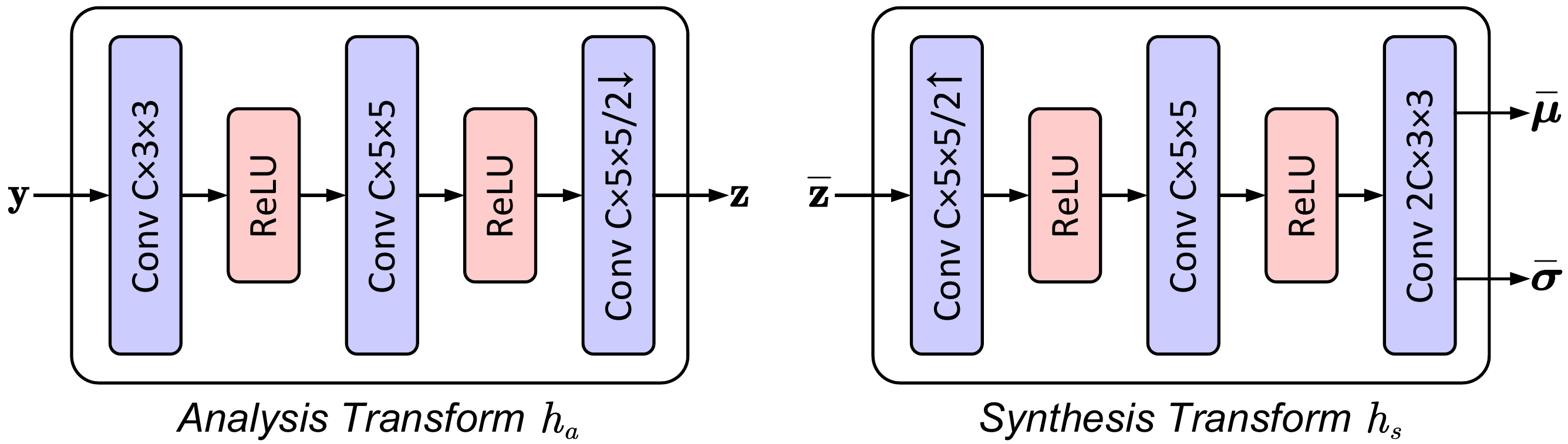}}
	\caption{Network architectures of nonlinear transforms $h_a$ and $h_s$. Convolution parameters are denoted as: the number of filters $\times$ kernel height $\times$ kernel width / up-sampling or down-sampling stride, where $\uparrow / \downarrow$ indicates up-sampling and down-sampling, respectively.}\label{Fig6}
	\vspace{0em}
\end{figure}

\begin{figure*}[t]
	\setlength{\abovecaptionskip}{0.cm}
	\setlength{\belowcaptionskip}{-0.cm}
	\centering{\includegraphics[scale=0.33]{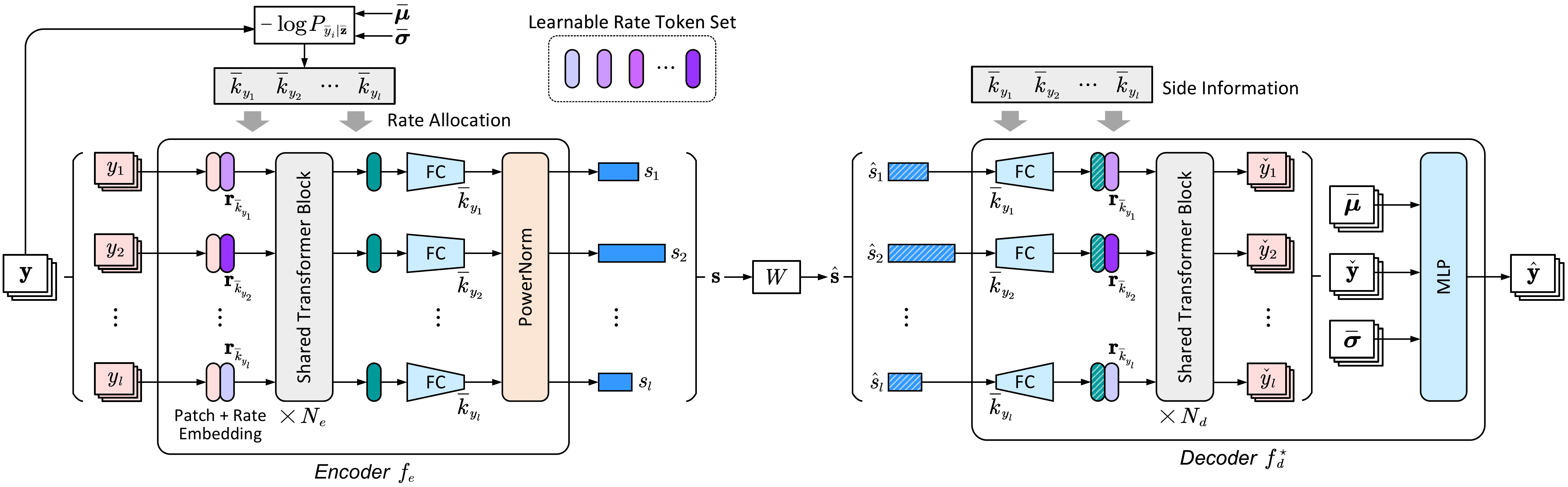}}
	\caption{Network architectures of codec $f_e$ and $f_d^{\star}$.}\label{Fig7}
	\vspace{0em}
\end{figure*}

\begin{figure*}[t]
	\setlength{\abovecaptionskip}{0.cm}
	\setlength{\belowcaptionskip}{-0.cm}
 \begin{center}
 \hspace{-.0in}
 \subfigure[]{
   \includegraphics[scale=0.17]{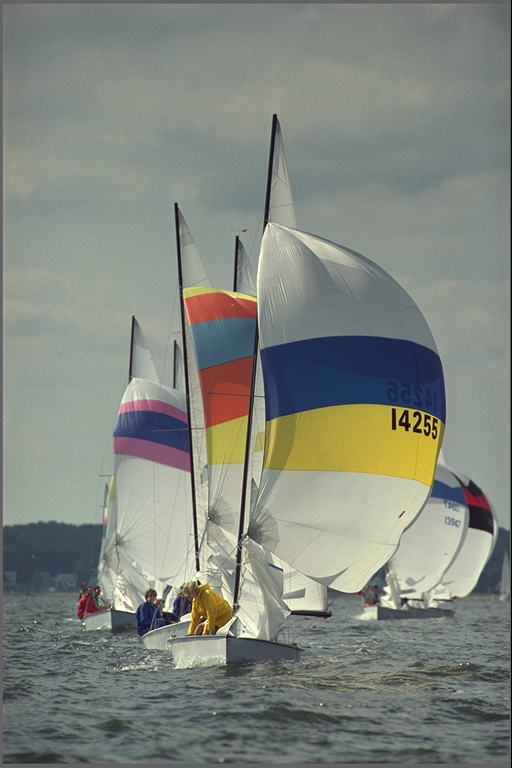}
 }
 \hspace{.10in}
 \quad
    \subfigure[]{
 \includegraphics[scale=0.31]{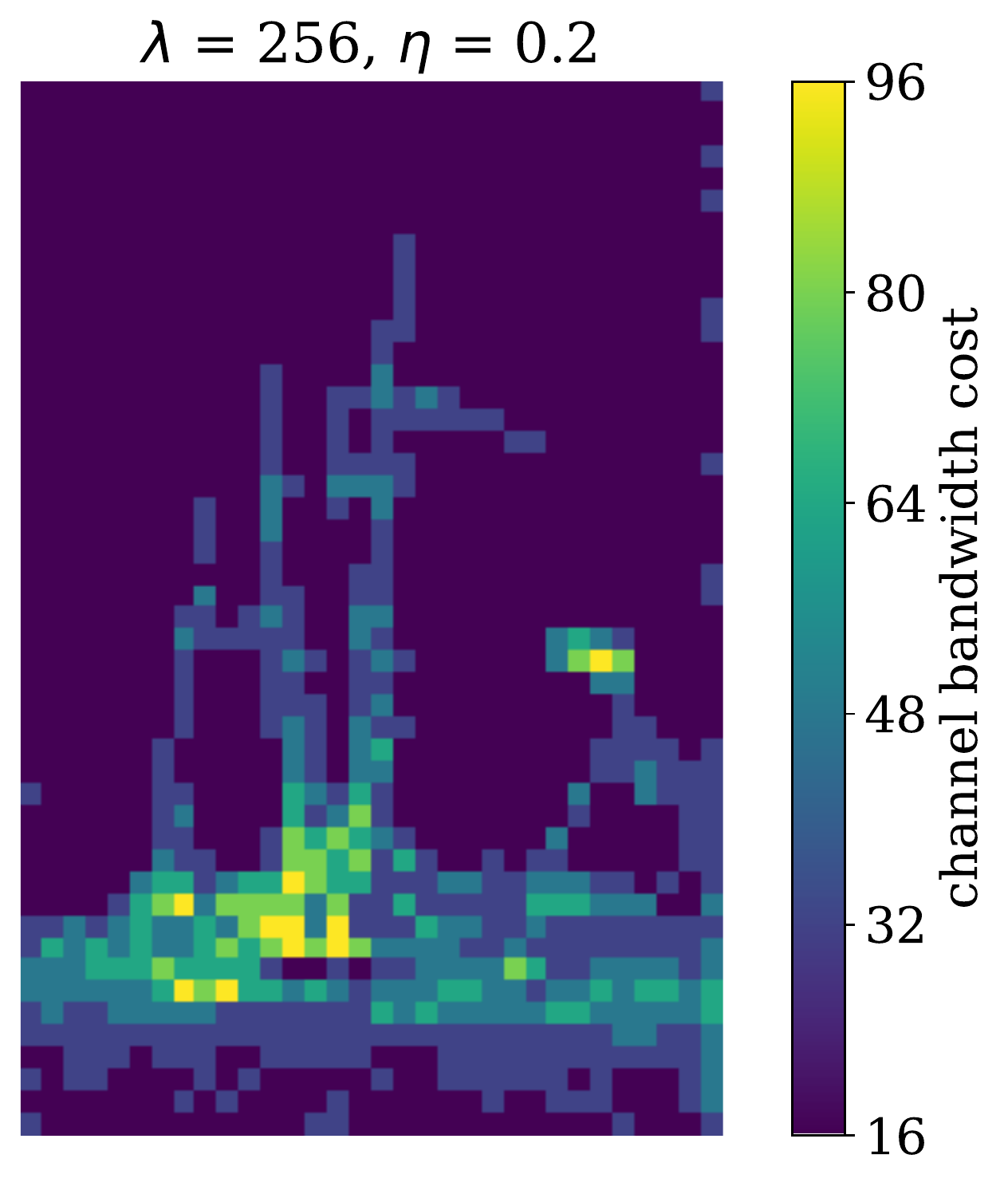}
 }
  \hspace{-.10in}
 \quad
    \subfigure[]{
 \includegraphics[scale=0.31]{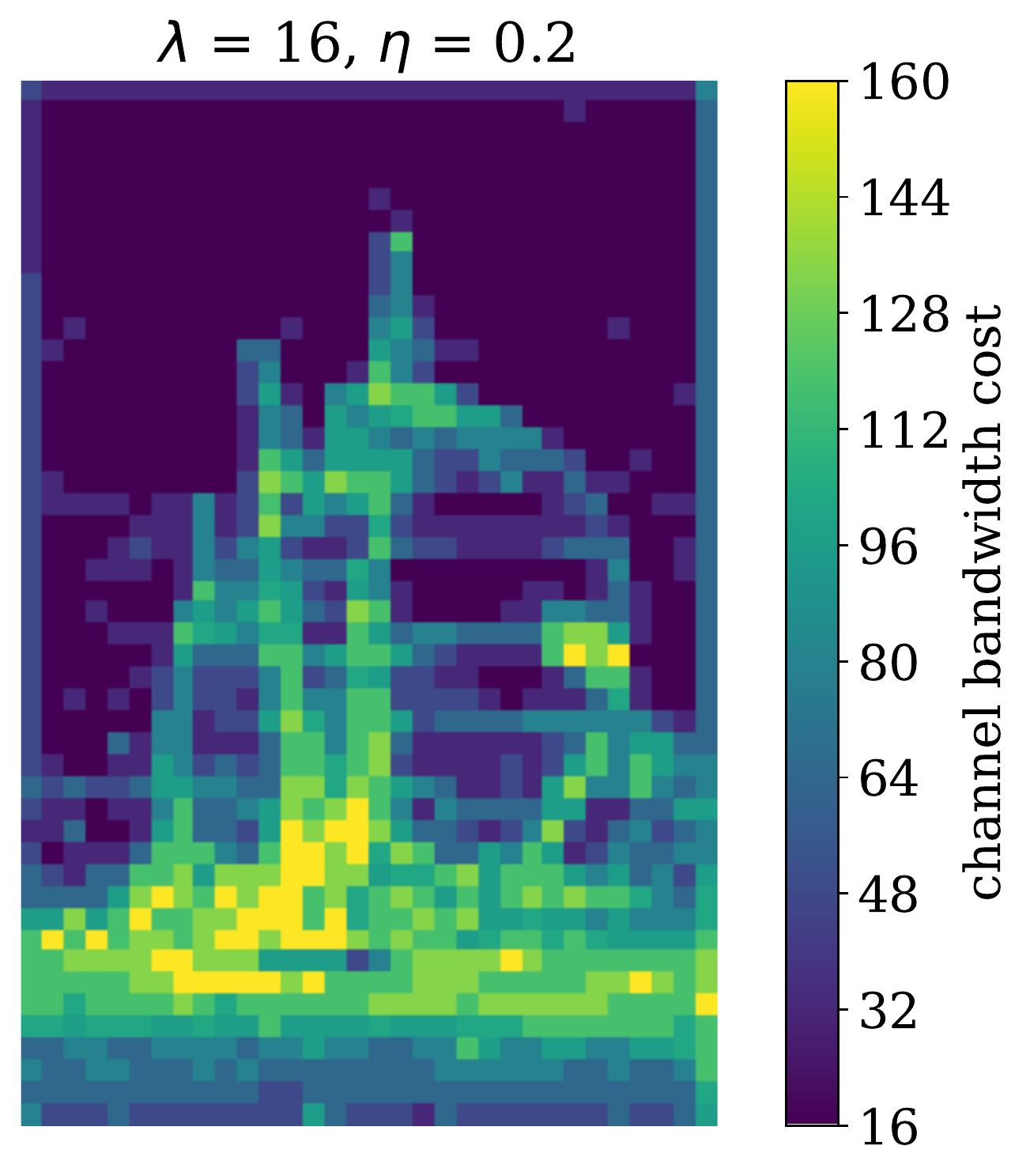}
 }
  \hspace{-.10in}
 \quad
    \subfigure[]{
 \includegraphics[scale=0.31]{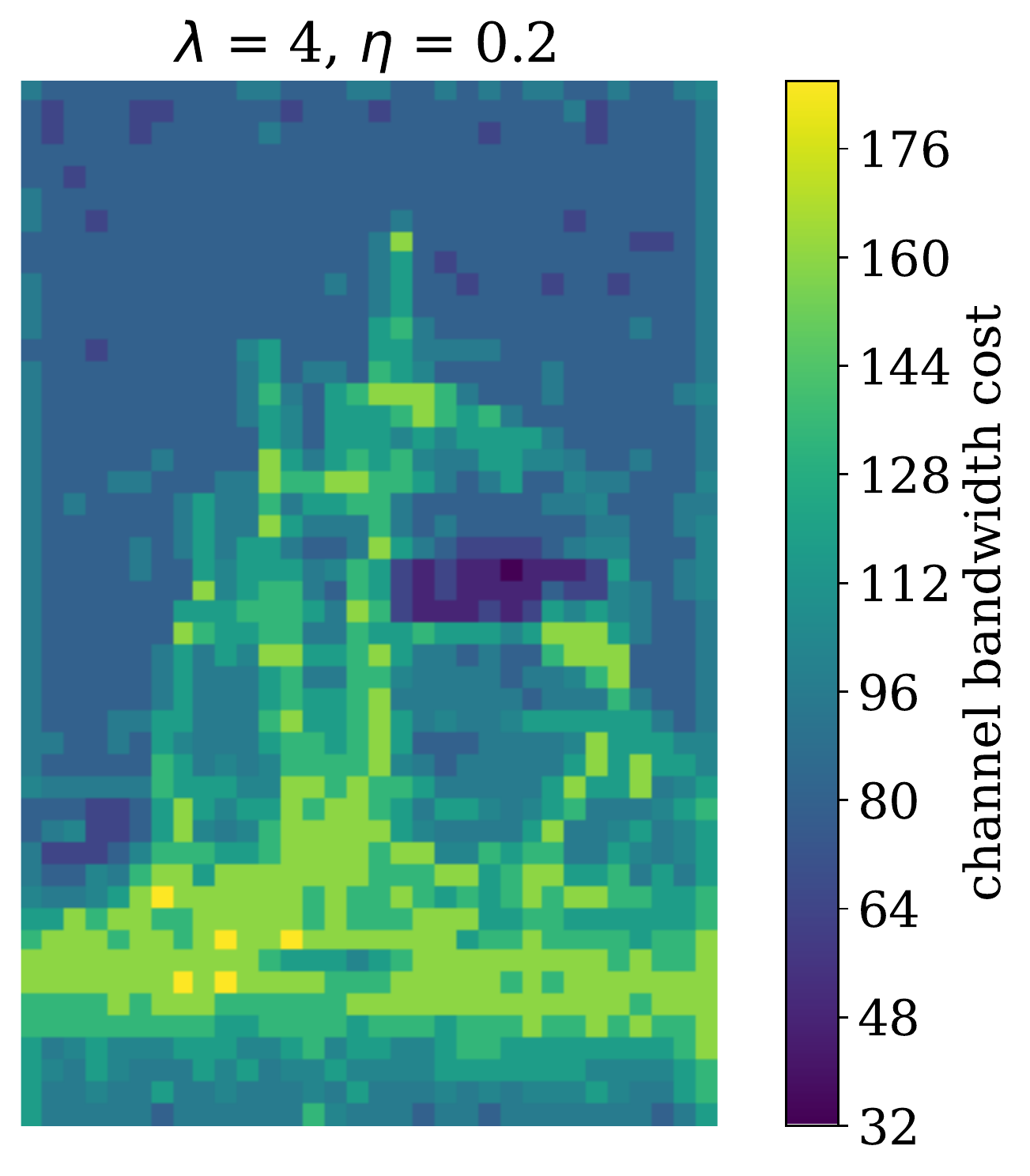}
 }
 \caption{A toy visualization demo of the rate adaption results. Apart from the original image, the other three figures demonstrate the allocated channel bandwidth cost for each patch, which are adjusted by the hyperparameters $\lambda$ and $\eta$.}
 \label{Fig8}
 \end{center}
\end{figure*}

After that, the analysis transform $g_a$ merges neighboring embeddings and reduces the concatenated $4c$-dimensional embeddings to $c$-dimensional by a linear FC layer, which achieves down-sampling of resolution by two times. Then, it feeds the $l_2=\frac{h}{4} \times \frac{w}{4}$ higher-resolution patch embedding tokens to another $N_2$ Transformer blocks. As shown in Fig. \ref{Fig5}, we encapsulate a down-sampling layer and the following Transformer blocks as one stage, the aforementioned procedure has included two stages. For small images, e.g., the CIFAR10 dataset of $32\times32$ pixels per image \cite{CIFAR10}, $g_a$ with two stages is enough, thus the analysis transform module outputs the latent representation $\mathbf{y} \in \mathbb{R}^{\frac{h}{4} \times \frac{w}{4} \times c}$. For large images, e.g., the CLIC2021 image dataset with size up to $2048\times 1890$ pixels \cite{CLIC2021}, one needs more stages to obtain higher-resolution patch embeddings, in this paper, we adopt $g_a$ with four stages as the analysis transform to produce the latent representation $\mathbf{y} \in \mathbb{R}^{\frac{h}{16} \times \frac{w}{16} \times c}$ as depicted in Fig. \ref{Fig5} for medium to large images.

The synthesis transform module $g_s$ has a symmetric architecture with analysis transform $g_a$, including the patch division operation for up-sampling and Transformers. It recovers input images from noisy or quantized latent representations. The hyperprior autoencoder, including $h_a$ and $h_s$, summarizes the distribution of means and standard deviations in $\mathbf{z}$. It consists of fully convolutional layers followed by the ReLU activation functions as shown in Fig. \ref{Fig6}.

\subsubsection{Codec Modules $f_e$ and $f_d^{\star}$}

The overall architectures of deep JSCC codec modules are illustrated in Fig. \ref{Fig7}. $f_e$ learns to transmit the latent representation $\mathbf{y}$ with variable transmission rate in accordance to the entropy model $P_{\mathbf{\bar y}|\mathbf{\bar z}}$. In particular, the encoder $f_e$ first partitions the latent representation $\mathbf{y}$ into patch embedding sequence $(y_1,y_2,\dots,y_l)$. As we shall note, each $y_i$ is a $c$-dimensional feature vector. The learned entropy model $-\log P_{{\bar y}_i|\mathbf{\bar z}}({\bar y}_i|\mathbf{\bar z})$ indicates the summation of entropy along all $c$ dimensions of $y_i$, thus, the information density distribution of $\mathbf{y}$ has been well captured. Accordingly, the channel bandwidth cost ${\bar k}_{y_i}$ for transmitting $y_i$ can be determined as
\begin{equation}\label{eq_channel_bandwidth_cost_cal}
\begin{aligned}
  {\bar k}_{y_i} & = Q'({ k}_{y_i})\\
   ~ & = Q'(-\eta\log P_{{\bar y}_i|\mathbf{\bar z}}({\bar y}_i|\mathbf{\bar z})) = Q'(-\eta\log p_{{\tilde y}_i|\mathbf{\tilde z}}({\bar y}_i|\mathbf{\bar z})),
\end{aligned}
\end{equation}
where the learned entropy model $p_{{\tilde y}_i|\mathbf{\tilde z}}$ follows \eqref{eq_ntscc_z_to_y_factorized}, $Q'$ denotes a scalar quantization whose range includes $2^{k_q}$ ($k_q = 1,2,\dots$) integers, and the quantization value set $\mathcal{V} = \{ v_1, v_2, \dots, v_{2^{k_q}} \}$ is related to the scaling factor $\eta$ and the Lagrange multiplier $\lambda$. Hence, a predetermined $k_q$ bits must be transmitted from Alice to Bob as an extra side information to inform the receiver which rate is allocated to each $y_i$.

As a toy visualization demo, Fig. \ref{Fig8} shows the rate adaption results of the left top image, where each patch is assigned a channel bandwidth value within $\mathcal{V}$. It can be observed that complex regions (water, text, and human) are encoded with more channel bandwidth costs, while simple areas (sky and canvas) use fewer. Furthermore, with the decrease of $\lambda$, the NTSCC model tends to learn the latent representation with higher entropy to reduce the distortion. As a result, the learned entropy model guides the $f_e$ to use more channel bandwidth for transmission.

To adaptively map $y_i$ to a ${\bar k}_{y_i}$-dimensional channel-input vector $s_i$, an intuitive solution is dividing the encoder $f_e$ as $l$ parts, each one is responsible to encode a patch embedding $y_i$. To attain this, one needs to train $2^{k_q}$ pairs of sub-encoders and sub-decoders. Apparently, this tough method will lead to large training complexity and storage costs. Furthermore, it ignores the contextual dependencies among $y_i$, which will degrade the system performance. Therefore, in this paper, we introduce the dynamic neural network structure \cite{han2021dynamic} into Transformers to realize the deep JSCC encoder $f_e$. Compared to static models of fixed computational graphs and parameters, dynamic networks can adapt their network structures or parameters to different inputs, leading to notable advantages in terms of performance, computational efficiency, etc \cite{han2021dynamic}. Here, we employ $N_e$ shared Transformer blocks for feature extraction and light FC layers to encode $y_i$ into given dimension ${\bar k}_{y_i}$. The process includes two steps:
\begin{enumerate}[(a)]
  \item Indicating the shared Transformer blocks of the channel bandwidth cost ${\bar k}_{y_i}$ of each $y_i$ through self-attention;
  \item Using light FC layers to transform the Transformer output corresponding to $y_i$ into a ${\bar k}_{y_i}$-dimensional vector.
\end{enumerate}

Inspired by positional embedding of Transformer, which enables the network to distinguish the relative positions of patches through position-related bias \cite{ViT}, we develop a rate token vector set $\mathcal{R} = \{ \mathbf{r}_{v_1}, \mathbf{r}_{v_2}, \dots, \mathbf{r}_{v_{2^{k_q}}} \}$ to indicate the rate information. As shown in Fig. \ref{Fig7}, each $y_i$ will be added its corresponding rate token vector $\mathbf{r}_{{\bar k}_{y_i}}$ before fed into Transformer blocks. As a result, the output of shared Transformers can learn to adapt to the entropy of $y_i$, and the following FC layer can efficiently adjust output dimension.

In the receiver, noisy channel vector $\hat{s}_i$ with different length will be reshaped to the unified dimension by FC blocks, then added the rate tokens and sent into $N_d$ shared Transformer blocks in parallel. The tentatively recovered version of $y_i$ is $\check{y}_i$. As stated in \eqref{eq_ntscc_term4_2}, the deep JSCC decoder can further refine the reconstructions $\check{\mathbf{y}}$ with the help of the learned prior $\bm{\bar \mu}$ and $\bm{\bar \sigma}$ that is realized by a two-layer MLP network in $f_d^{\star}$.

\section{Experimental Results}\label{section_performance}

\subsection{Experimental Setup}

\subsubsection{Datasets}

We quantify the wireless image transmission performance on RGB small-size image dataset CIFAR10 \cite{CIFAR10} (50,000 training images and another 10,000 test images with $32\times 32$ pixels), medium-size dataset Kodak \cite{Kodak} (24 images, $768\times 512$ pixels), and large-size dataset CLIC2021 \cite{CLIC2021} (60 images, up to $2048\times 1890$ pixels). The content and resolution of these datasets are diversified, thus have been widely used to measure the performance of image-related algorithms. In addition, the dataset for training the proposed NTSCC model for large images consists of 500,000 images sampled from the Open Images Dataset \cite{OpenImage}. During model training, images are randomly cropped into $256\times 256$ patches.

\subsubsection{Comparison Schemes}

Our comparison schemes include both the emerging deep JSCC scheme proposed in \cite{DJSCC} and the classical separation-based source and channel coding schemes. Specifically, we employ the powerful image codec BPG \cite{BPG} combined with a practical LDPC code \cite{LDPC_5G} or an ideal capacity-achieving channel code for the latter scheme (marked as ``BPG + LDPC'' and ``BPG + Capacity'', respectively). As we shall note, the ideal ``BPG + Capacity'' scheme can be viewed as a performance bound on the separation-based source and channel coding schemes. In practical implementation, to be aligned with previous works \cite{DJSCC}, we convert two consecutive real symbols in $\mathbf{s}$ as one complex channel-input symbol and add complex Gaussian noise.

\subsubsection{Evaluation Metrics}

We qualify the end-to-end image transmission performance of the proposed NTSCC models and state-of-the-art models using both the widely used pixel-wise metric (i.e., the peak signal-to-noise ratio (PSNR)) and the perceptual metric (i.e., the multi-scale structural similarity index (MS-SSIM) \cite{msssim}) and the emerging deep learning based perceptual metric (i.e., the learned perceptual image patch similarity (LPIPS) \cite{lpips}). PSNR corresponds to the pixel-wise $\ell_2$ Euclidian distance, thus when evaluating PSNR performance, we set the distortion function $d$ as the mean square error (MSE) between $\mathbf{x}$ and $\mathbf{\hat x}$. When evaluating the MS-SSIM performance, the training loss $d$ is set as ``$1-\text{MS-SSIM}$'' so as to minimize $d$. It is usually known that a higher PSNR/MS-SSIM indicates a better performance.

Although PSNR and MS-SSIM are the most widely used metrics, they are simple and shallow functions, and fail to account for many nuances of human perception \cite{lpips}. To be more closely aligned with the purpose of semantic communications, we further adopt the emerging deep learning based LPIPS metric \cite{lpips} as a perceptual loss to quantify the image transmission performance, it can imitate the human perceptual assessment process to give the LPIPS loss score. A lower LPIPS value indicates a lower distortion, with a maximum value of 1. In practical model training, one can observe that the weakness of each metric is exploited by learning algorithms, e.g., checkerboard artifacts may appear when targeting neural network derived by the LPIPS metric, relying on MS-SSIM can cause poor text reconstructions, and MSE yields blurry reconstructions \cite{HIFIC,ding2021comparison}. To tackle this, when the NTSCC model targets at human subjective perception metric, we introduce a conditional generative adversarial network (cGAN) \cite{mirza2014conditional} and augment the MSE distortion item with GAN loss and LPIPS loss. The total distortion item for model training is
\begin{equation}\label{eq_lpips_loss}
\begin{aligned}
d(\mathbf{x},\mathbf{\hat{x}}_{\text{NTSCC}}) = & \beta_M \| \mathbf{x} -{\mathbf{\hat{x}}}_{\text{NTSCC}} \|^2 + \beta_L d_{\text{LPIPS}}(\mathbf{x},{\mathbf{\hat{x}}}_{\text{NTSCC}})- \\
~ & \beta_D \log(D({\mathbf{\hat{x}}}_{\text{NTSCC}}, \mathbf{\tilde{y}})),
\end{aligned}
\end{equation}
where $\beta_M, \beta_L, \beta_D$ control the trade-off between three distortion terms, $d_{\text{LPIPS}}$ is the LPIPS distortion, and $D$ denotes the discriminator of the cGAN, which maps the latent representation and reconstruction to the probability that it is a sample from the true distribution $p_{\mathbf{x}|\mathbf{y}}$ rather than from the synthesis transform. Here, the discriminator adopts a fully convolutional network with the same architecture to \cite{HIFIC}.

\subsubsection{Model Training Details}

\begin{figure*}[t]
\setlength{\abovecaptionskip}{0.cm}
\setlength{\belowcaptionskip}{-0.cm}
\begin{center}
	\hspace{-.10in}
	\subfigure[]{
		\includegraphics[scale=0.4]{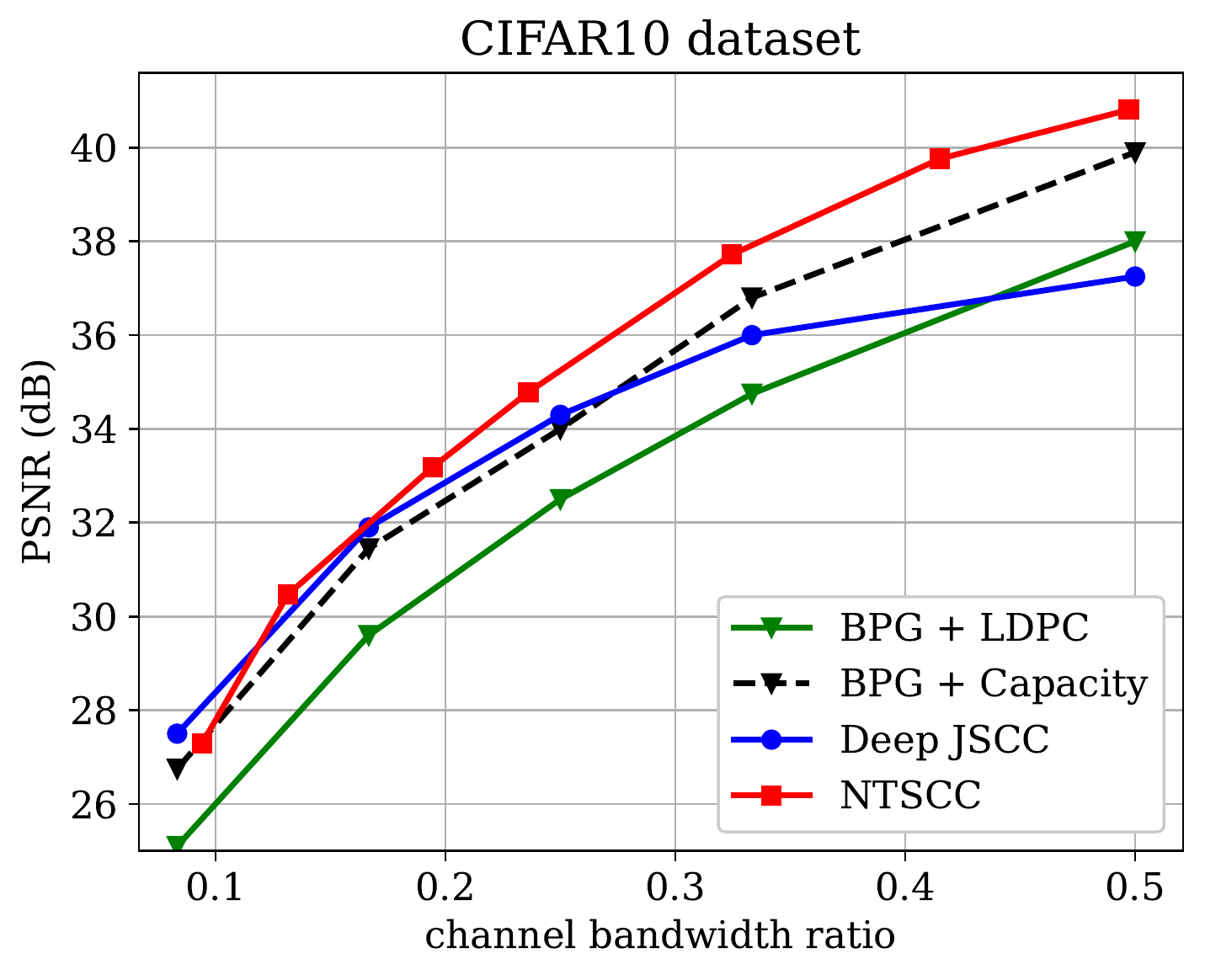}
	}
	\hspace{-.30in}
	\quad
	\subfigure[]{
		\includegraphics[scale=0.4]{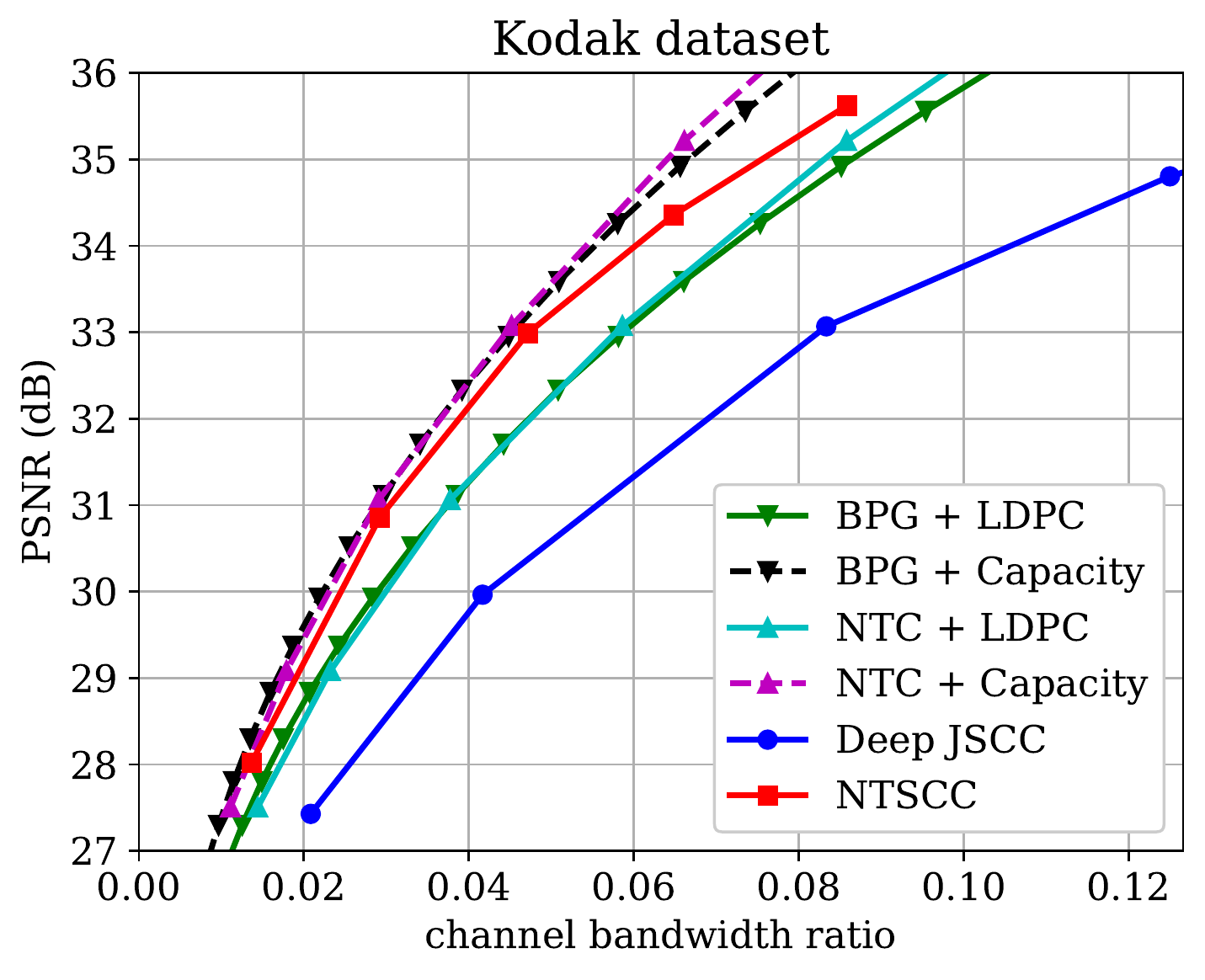}
	}
	\hspace{-.30in}
	\quad
	\subfigure[]{
        \includegraphics[scale=0.4]{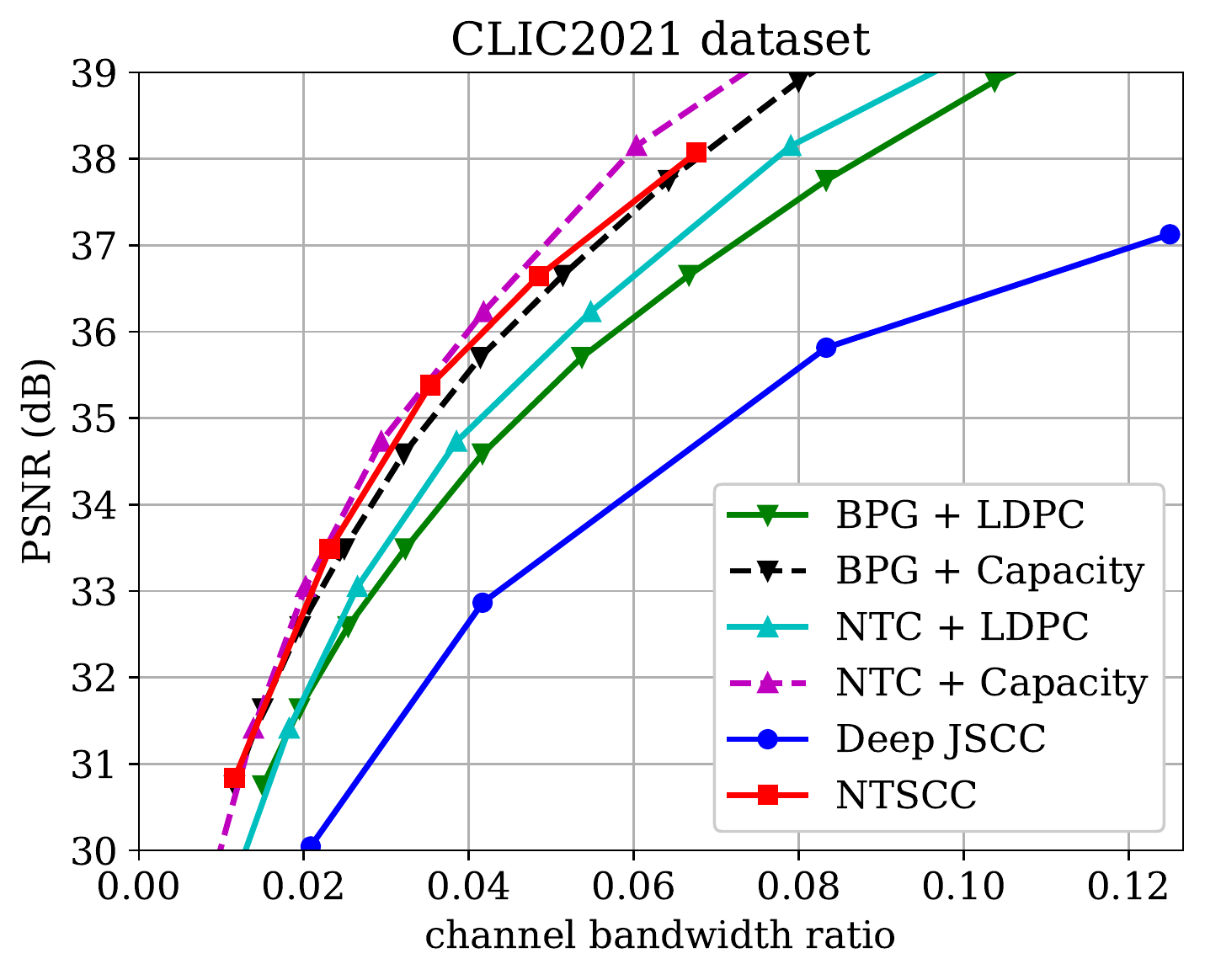}
	}
	\caption{PSNR performance versus the average channel bandwidth ratio (CBR) over the AWGN channel at $\text{SNR} = 10$dB.}
		\label{Fig9}
\end{center}
\end{figure*}

As aforementioned, we set up the analysis transform $g_a$ and the synthesis transform $g_s$ with $4$ stages with $N_1 = 1, N_2 = 1, N_3 = 2, N_4 = 6$ Transformer blocks for Kodak and CLIC2021 datasets while using $2$ stages with $N_1 = 2, N_2 = 6$ Transformer blocks for the tiny CIFAR10 dataset. In all experiments, we use $N_e = N_d = 4$ transformer blocks in JSCC codec $f_e$ and $f_d^{\star}$, the number of heads in MHSA is $8$, and the channel dimension $c$ is set to $256$. The quantization value set $\mathcal{V}$ consists of 16 values, which are evenly distributed from 16 to 256. Therefore, an extra side information of $k_q = 4$ bits will be transmitted to inform the receiver which channel bandwidth cost is allocated for each patch. Besides, the complexity of the self-attention calculation is quadratic to the length of the token sequence, resulting in unaffordable computation/memory cost for the transmission of high-resolution images (e.g., $2048\times 1890$ pixels of CLIC2021 dataset). To tackle this, we employ the shifted-windows-based self-attention mechanism (Swin Transformer) in \cite{liu2021Swin} for large-scale images, which significantly reduces the computational and memory overhead by computing MHSA within non-overlapped local windows size of $16 \times 16$.

During the training process, we use the Adam optimizer \cite{ADAM} with a learning rate of $10^{-4}$. The mini-batch size is set as 10. All implementations were done on Pytorch \cite{paszke2019pytorch}, and it takes about 4 days from pretrained NTC models to train the whole NTSCC model using single RTX 3090 GPU.

\subsection{Results Analysis}

\begin{figure*}[t]
\begin{center}
\setlength{\abovecaptionskip}{0.cm}
\setlength{\belowcaptionskip}{-0.cm}
	\hspace{-.10in}
	\subfigure[]{
		\includegraphics[scale=0.4]{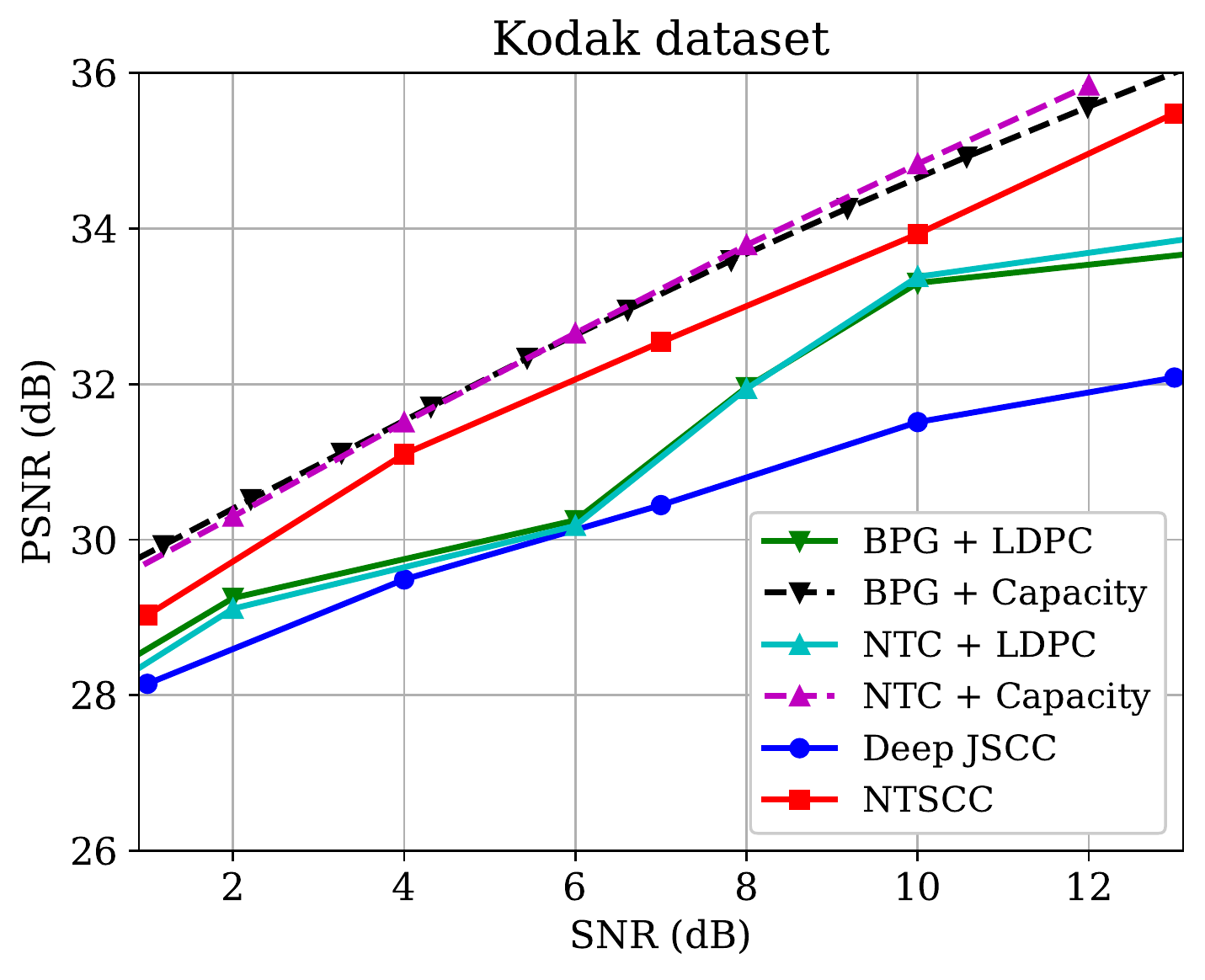}
	}
	\hspace{-.30in}
	\quad
	\subfigure[]{
	   \includegraphics[scale=0.4]{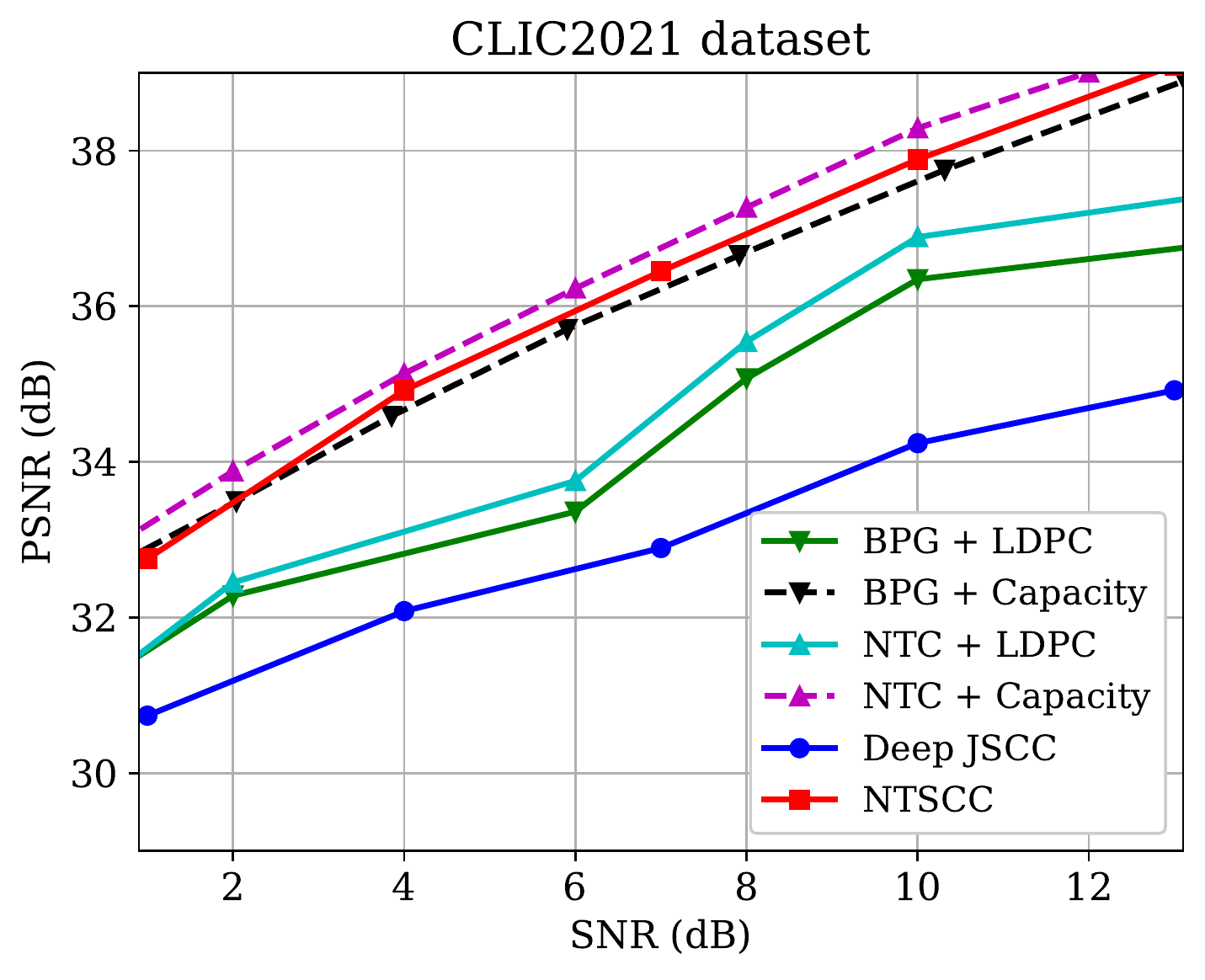}
	}
    \hspace{-.30in}
	\quad
	\subfigure[]{
        \includegraphics[scale=0.4]{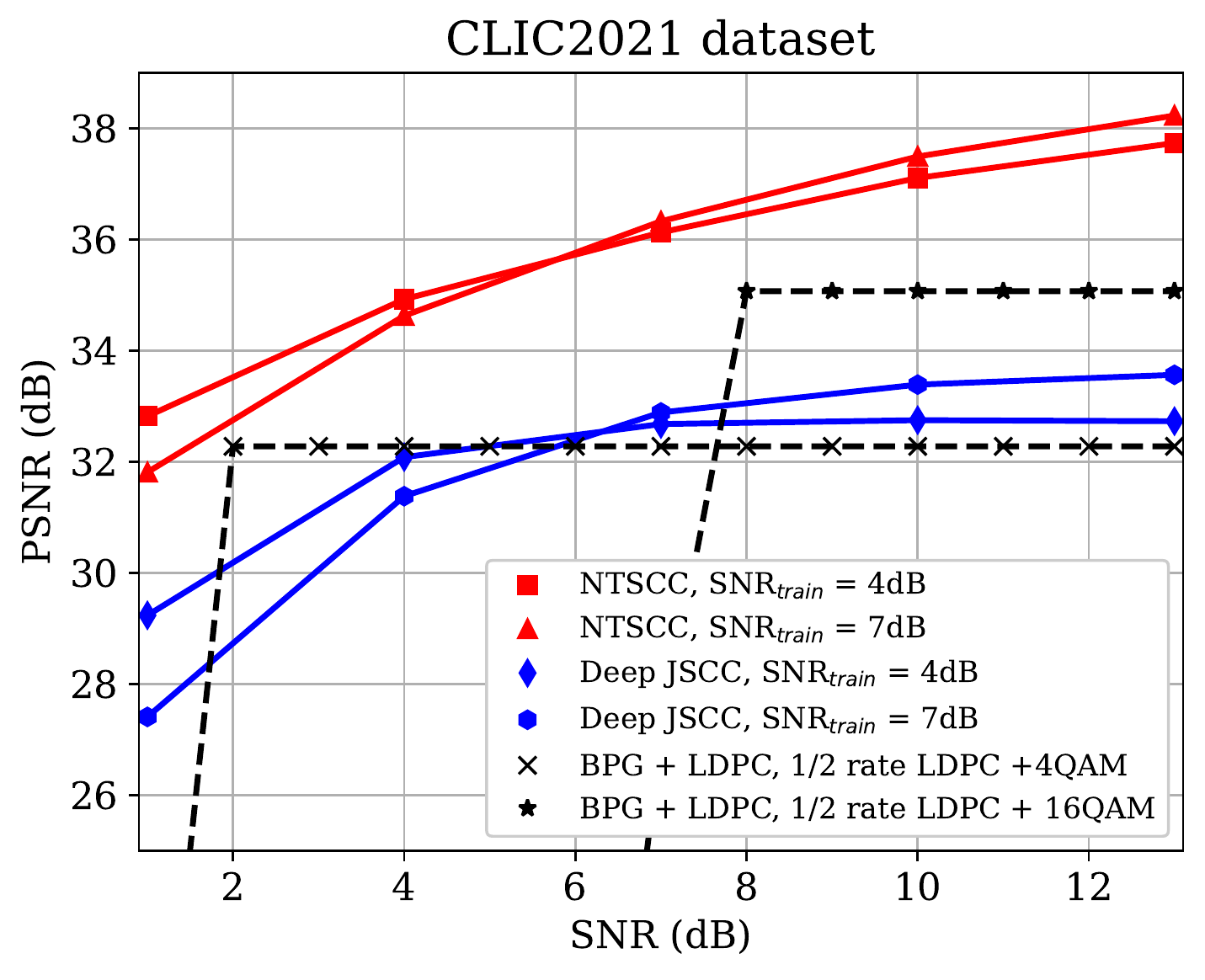}
	}
	\caption{PSNR performance versus the SNR at the AWGN channel for medium to high resolution image datasets, and the average CBR is set to $R = k/m = 1/16$. As we shall note, for the NTSCC scheme, its CBR is not easy to be settled down to a predetermined value, thus, to ensure a fair comparison, the maximum $R$ in NTSCC is constrained to $1/16$.}
	\label{Fig10}
\end{center}
\end{figure*}

\begin{table*}[t]
\renewcommand{\arraystretch}{1.3}
  \centering
  \small

  \caption{The BD-CBR and BD-PSNR performances comparison.}

  \begin{tabular}{!{\vrule width1pt}m{2.2cm}!{\vrule width1pt}m{1.8cm}|m{2.2cm}!{\vrule width1pt}m{1.8cm}|m{2.2cm}!{\vrule width1pt}m{1.8cm}|m{2.2cm}!{\vrule width1pt}}

    \Xhline{1pt}

     \centering \multirow{2}*{{Method}} & \multicolumn{2}{c!{\vrule width1pt}}{CIFAR10 dataset} & \multicolumn{2}{c!{\vrule width1pt}}{Kodak dataset} & \multicolumn{2}{c!{\vrule width1pt}}{CLIC2021 dataset} \tabularnewline

    \cline{2-7}
     & \centering BD-CBR &  \centering BD-PSNR (dB) & \centering BD-CBR & \centering BD-PSNR (dB) & \centering BD-CBR & \centering BD-PSNR (dB) \tabularnewline

     \hline
     \centering NTSCC & \centering {\textbf{--28.91\%}} & \centering {\textbf{2.64}} & \centering --17.96\% & \centering 0.81 & \centering {\textbf{--28.09\%}} & \centering {\textbf{1.31}} \tabularnewline

 \hline
     \centering BPG + Capacity & \centering --21.41\% & \centering 1.77 & \centering {\textbf{--22.91\%}} & \centering{\textbf{0.97}} & \centering --22.91\% & \centering 1.09 \tabularnewline

 \hline
     \centering BPG + LDPC & \centering 0.00\% & \centering 0 & \centering 0.00\% & \centering 0 & \centering 0.00\% & \centering 0 \tabularnewline

 \hline
     \centering Deep JSCC & \centering --22.71\% & \centering 1.71 & \centering 45.48\% & \centering --1.55 & \centering 54.31\% & \centering --1.76 \tabularnewline

    \Xhline{1pt}
  \end{tabular}
  \label{Table1}
\end{table*}

\subsubsection{PSNR Performance}

Fig. \ref{Fig9} demonstrates the RD results among various transmission methods at the AWGN channel with $\text{channel SNR} = 10$dB, where the distortion is measured in terms of PSNR. To achieve different RD tradeoffs, we use NTSCC models trained with $\lambda \in \{  1024, 256, 64, 16, 4 \}$ and set $\eta = 0.2$. For separation schemes, we employ a $2/3$ rate (4096, 6144) LDPC code with 16-ary quadrature amplitude modulation (16QAM) for the ``BPG + LDPC'' scheme. The ideal ``BPG + Capacity'' scheme adopts the Gaussian channel capacity formula \cite{DJSCCF} as the transmission rate per channel bandwidth. From these results, we can find that the proposed NTSCC method can generally outperform deep JSCC for all CBRs, and their performance gap increases with the resolution of the image dataset and the channel bandwidth ratio. Compared with separation-based methods, the proposed NTSCC shows competitive performance to the ideal ``BPG + Capacity'' scheme and overpasses the practical ``BPG + LDPC'' for all three datasets. Furthermore, NTSCC can outperform ``NTC + LDPC'' which means the gains are not purely from better compression brought by NTC, and there is an element of good match between the learned deep JSCC and the nonlinear transform. In addition, the proposed NTSCC can closely approach the performance of ``NTC + Capacity'' which is an upperbound of NTC combined with channel coding. In fact, ``NTC + Capacity'' is only an approximated upperbound of NTSCC. The reason is that NTC suffers from some extra performance loss caused by the quantization of $\mathbf{y}$, while our NTSCC can directly exploit $\mathbf{y}$ for transmission, that may even bring some performance gain.

As we shall note, with the help of the adaptive rate transmission mechanism, the proposed NTSCC method provides comparable coding gain as that of the BPG series in most cases, while conventional deep JSCC degrades rapidly with the increase of bandwidth ratio. We also note that the NTSCC may not maintain this coding gain all the time, especially at very high bandwidth ratio regions. The phenomenon stems from the ANN structure of $g_a$, where we set the number of channels at the output layer as $c = 256$, thus the following deep JSCC encoder $f_e$ can only support ${\bar k}_{y_i} = 256$ dimensions for each patch embedding $y_i$ at most. Otherwise, it may need an extra FC layer to expand the output dimension, this operation will degrade the coding gain. As a result, many patches at the high bandwidth ratio region have been saturated to $256$ transmission dimensions. Besides, we also note that the NTSCC on the tiny CIFAR10 dataset performs even slightly worse than the deep JSCC scheme. The reason is that the coding gain brought by adaptive rate allocation and codec refinement may be taken away by the cost of transmitting side information, especially for low bandwidth ratio regions. However, the relative overhead for transmitting side information will become negligible for large-scale images.

Fig. \ref{Fig10}(a) and Fig. \ref{Fig10}(b) show the PSNR results of high-resolution images versus the change of channel SNR, where the average bandwidth ratio is $R = 1/16$. For the ``BPG + LDPC'' scheme, similar to \cite{DJSCCF}, we evaluate the performance of different combinations of LDPC coding rate and modulations and present the envelope of the best performing configurations at each SNR. Since the NTSCC method learns an adaptive coding rate allocation mechanism, we traverse the scaling factor $\eta \in [0.05,0.2]$ and finetune the NTSCC model with $\lambda = 16$ at each SNR to ensure the maximum CBR lower than $1/16$, thus achieving a fair comparison. Here, for all models, the training SNR equals the testing SNR. We find that the proposed NTSCC method brings considerable performance gain, outperforming the standard deep JSCC by a margin of at least 1dB. Furthermore, the coding gain of NTSCC holds with the increase of SNR. Like that in Fig. \ref{Fig9}, NTSCC can also outperform ``NTC + LDPC'' and closely approach the performance bound of ``NTC + Capacity''.

Besides, as observed in Fig. \ref{Fig10}(c), the proposed NTSCC also achieves graceful degradation as deep JSCC does when the testing SNR decreases from the training SNR, while the performance of separation-based ``BPG + LDPC'' transmission scheme reduces drastically (known as the \emph{cliff effect}).

Table \ref{Table1} provides BD-CBR and BD-PSNR results where the relative metric ``BD-X'' has been widely used to evaluate the performance of different image/video compression systems \cite{BDR}. Here, the BD-CBR represents the average percentage of CBR savings compared with the baseline scheme at the same PSNR. A negative number of BD-CBR stands for bandwidth saving, and a positive number means bandwidth cost increasing. The BD-PSNR represents the PSNR gains at the same average CBR. A positive number of BD-PSNR means image transmission quality improvement and vice versa. We view the ``BPG + LDPC'' as the baseline scheme. Results indicate that the proposed NTSCC method with two-stage $g_a$ can achieve 28.91\% bandwidth savings or PSNR gains of 2.64dB on the CIFAR10 dataset, and the NTSCC with four-stage $g_a$ can achieve  17.96\% and 28.09\% bandwidth savings or PSNR gains of 0.81dB and 1.31dB on the Kodak dataset and the CLIC2021 dataset, respectively. In fact, the most appropriate number of Transformer stages in the nonlinear transform module can be different for various source image resolutions, it will be explored in future. As analyzed before, deep JSCC performs worse on high-resolution datasets, thus there are 45.48\% and 54.31\% bandwidth costs increase.

\subsubsection{MS-SSIM Performance}

\begin{figure*}[t]
\setlength{\abovecaptionskip}{0.cm}
\setlength{\belowcaptionskip}{-0.cm}
	\begin{center}
		\hspace{-.10in}
		\subfigure[]{	\includegraphics[scale=0.3]{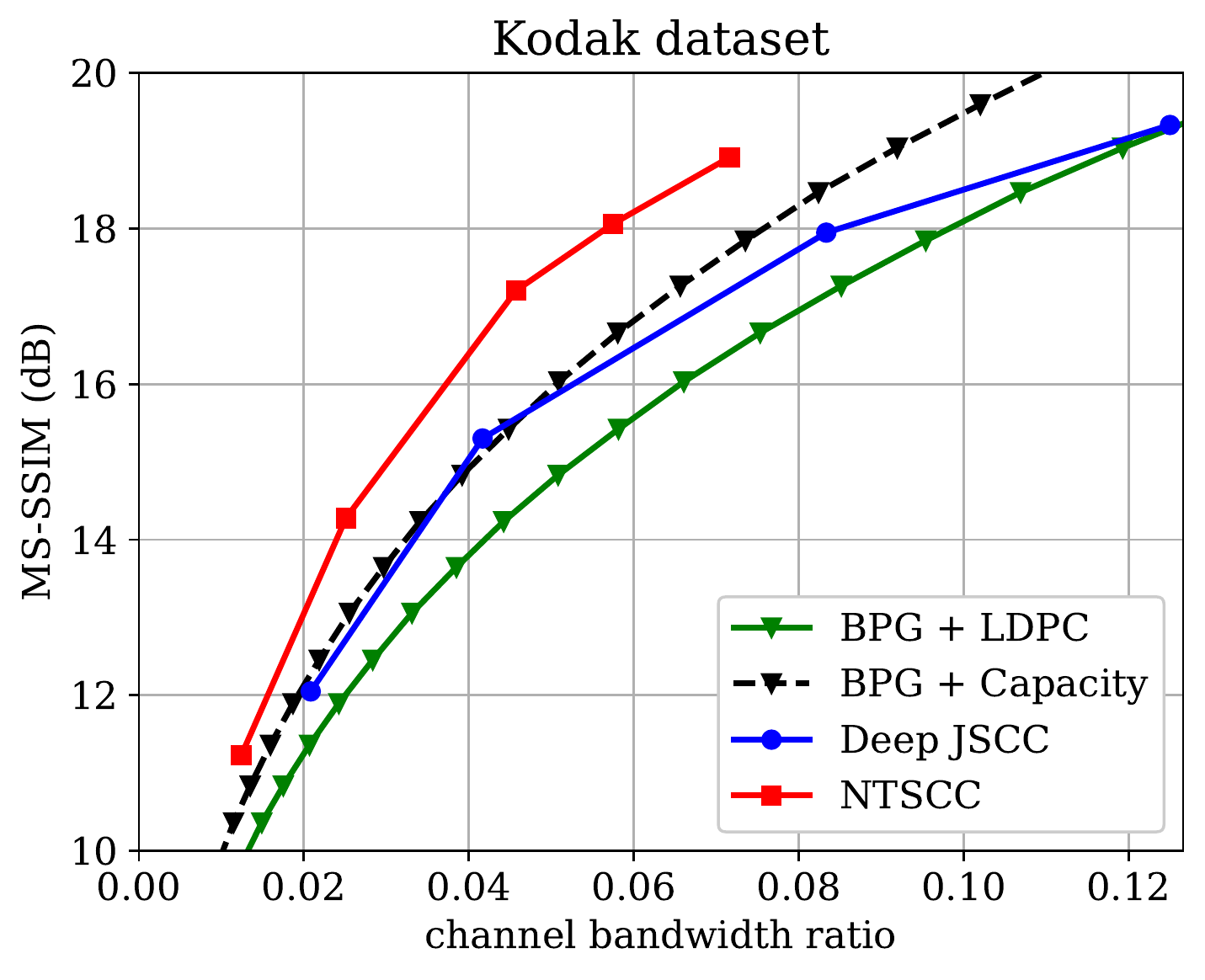}
		}
		\hspace{-.25in}
		\quad
		\subfigure[]{
			\includegraphics[scale=0.3]{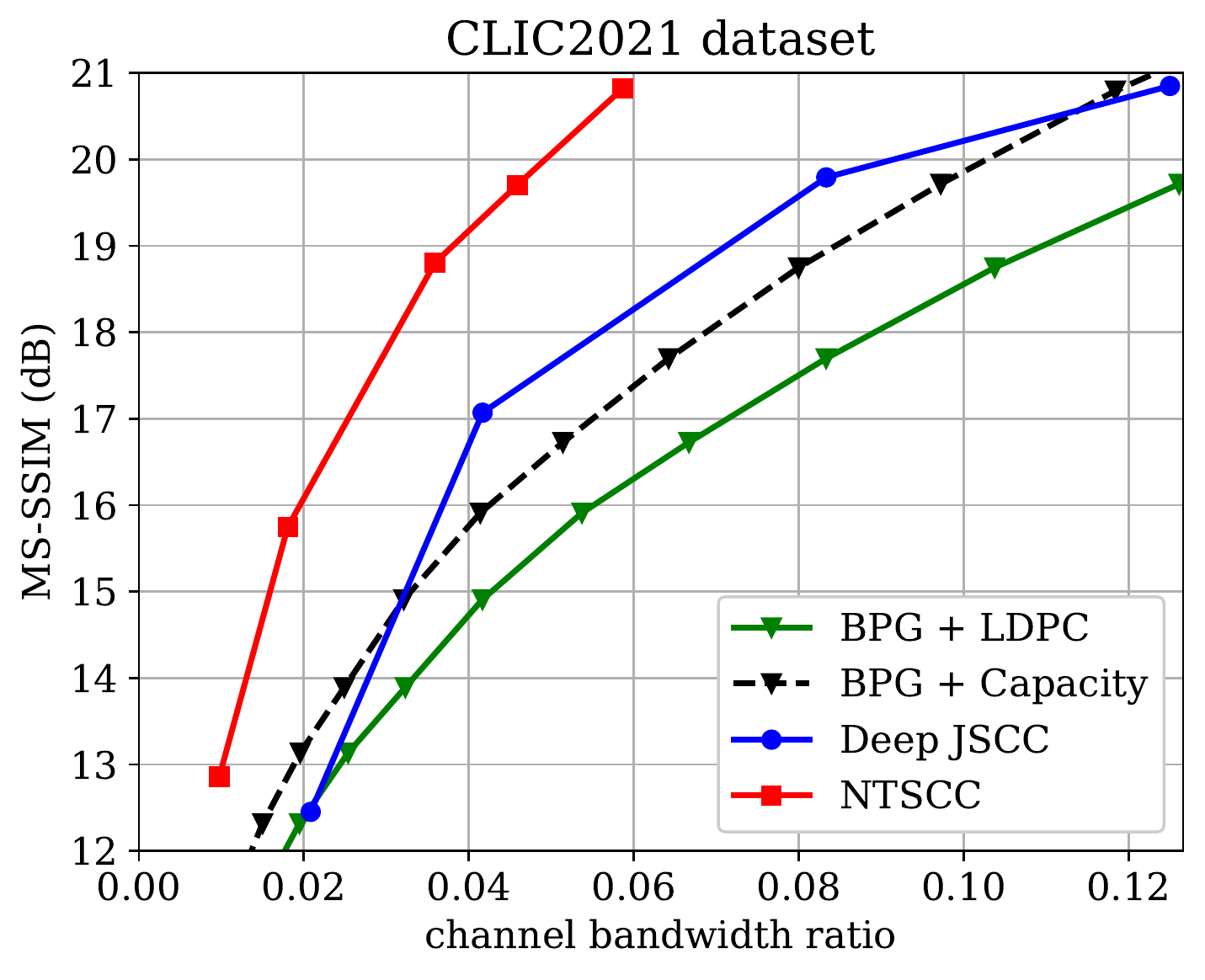}
		}
		\hspace{-.25in}
		\quad
		\subfigure[]{
			\includegraphics[scale=0.3]{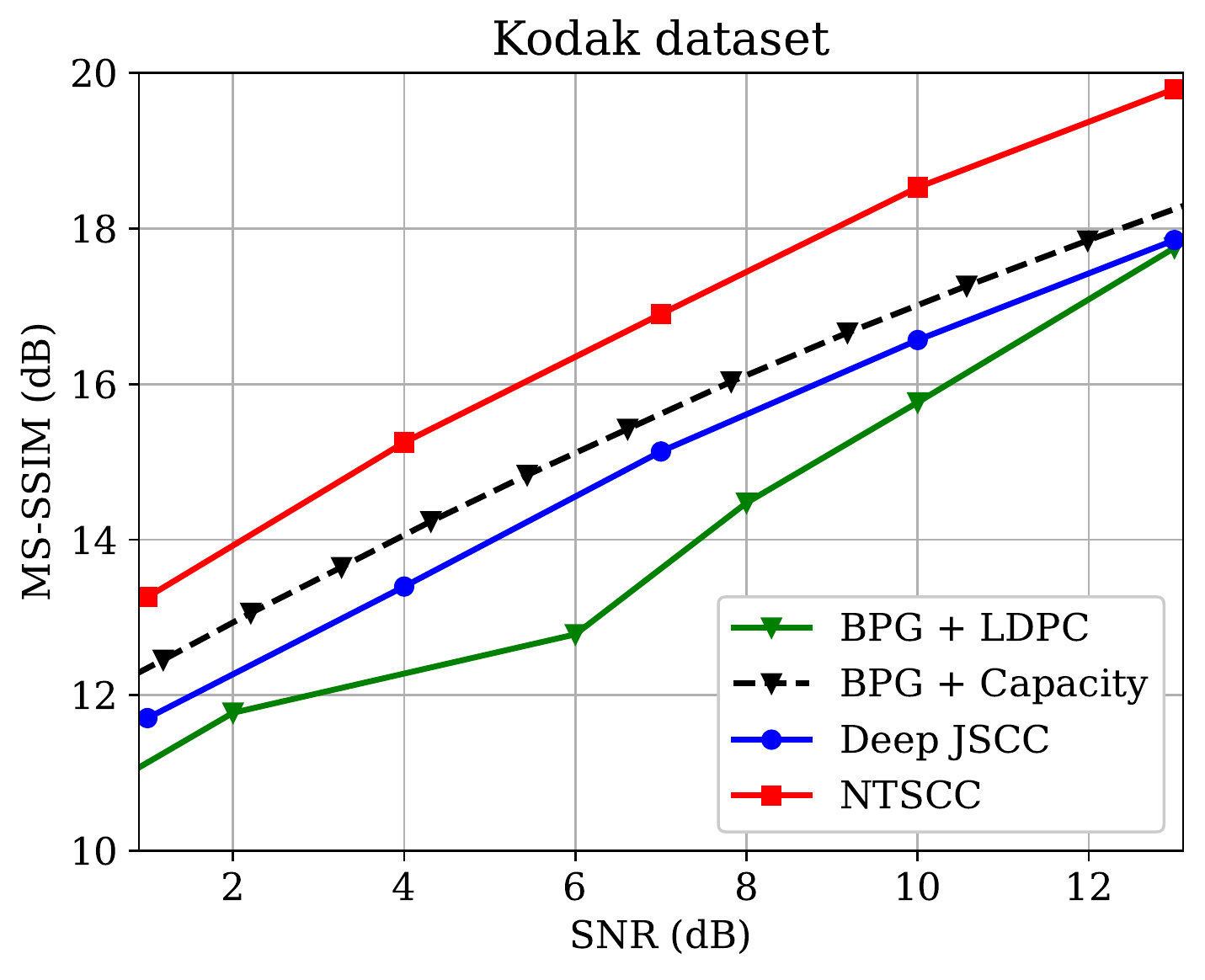}
		}
		\hspace{-.25in}
		\quad
		\subfigure[]{
			\includegraphics[scale=0.3]{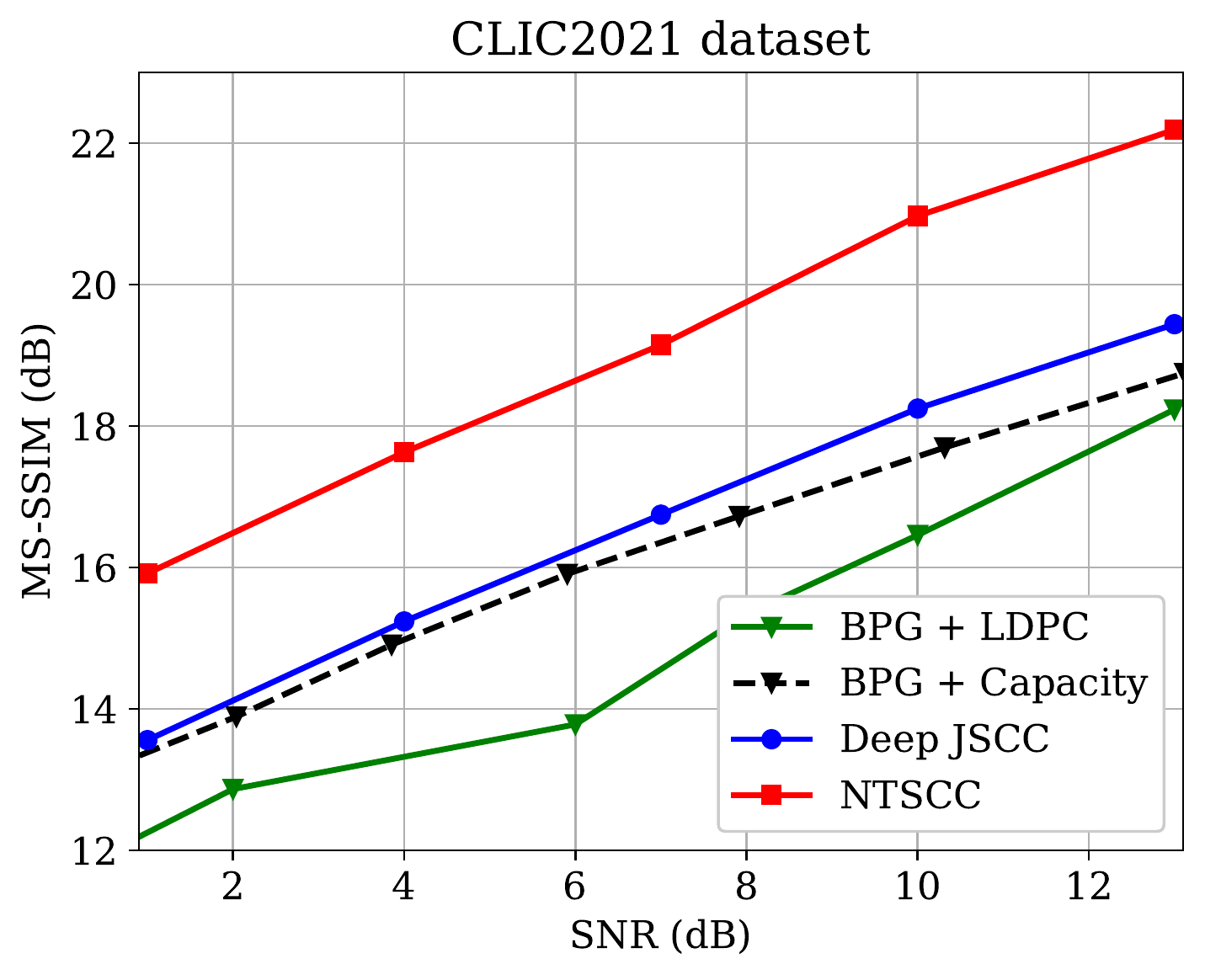}
		}
		\caption{(a) and (b) show the MS-SSIM performance versus the CBR over the AWGN channel at $\text{SNR} = 10$dB. (c) and (d) show the MS-SSIM performance versus the change of channel SNR, where the average CBR is set to $R = k/m = 1/16$.}
		\label{Fig11}
	\end{center}
\end{figure*}

\begin{table*}[t]
\renewcommand{\arraystretch}{1.3}
  \centering
  \small

  \caption{The BD-CBR and BD-MS-SSIM performances comparison.}

  \begin{tabular}{!{\vrule width1pt}m{2.2cm}!{\vrule width1pt}m{1.8cm}|m{3cm}!{\vrule width1pt}m{1.8cm}|m{3cm}!{\vrule width1pt}}

    \Xhline{1pt}

     \centering \multirow{2}*{{Method}} & \multicolumn{2}{c!{\vrule width1pt}}{Kodak dataset} & \multicolumn{2}{c!{\vrule width1pt}}{CLIC2021 dataset} \tabularnewline

    \cline{2-5}
     & \centering BD-CBR &  \centering BD-MS-SSIM (dB) & \centering BD-CBR & \centering BD-MS-SSIM (dB) \tabularnewline

     \hline
     \centering NTSCC & \centering {\textbf{--43.01\%}} & \centering {\textbf{2.29}} & \centering {\textbf{--64.27\%}} & \centering {\textbf{4.18}} \tabularnewline

     \hline
     \centering BPG + Capacity & \centering --20.99\% & \centering 1.30 & \centering --22.92\% & \centering 1.09 \tabularnewline

     \hline
     \centering BPG + LDPC & \centering 0.00\% & \centering 0 & \centering 0.00\% & \centering 0 \tabularnewline

     \hline
     \centering Deep JSCC & \centering --21.47\% & \centering  1.01 & \centering --31.77\% & \centering 1.77 \tabularnewline

    \Xhline{1pt}
  \end{tabular}
  \label{Table2}

\end{table*}

\begin{figure*}[t]
\setlength{\abovecaptionskip}{0.cm}
\setlength{\belowcaptionskip}{-0.cm}
	\begin{center}
		\hspace{-.05in}
		\subfigure[]{\includegraphics[scale=0.47]{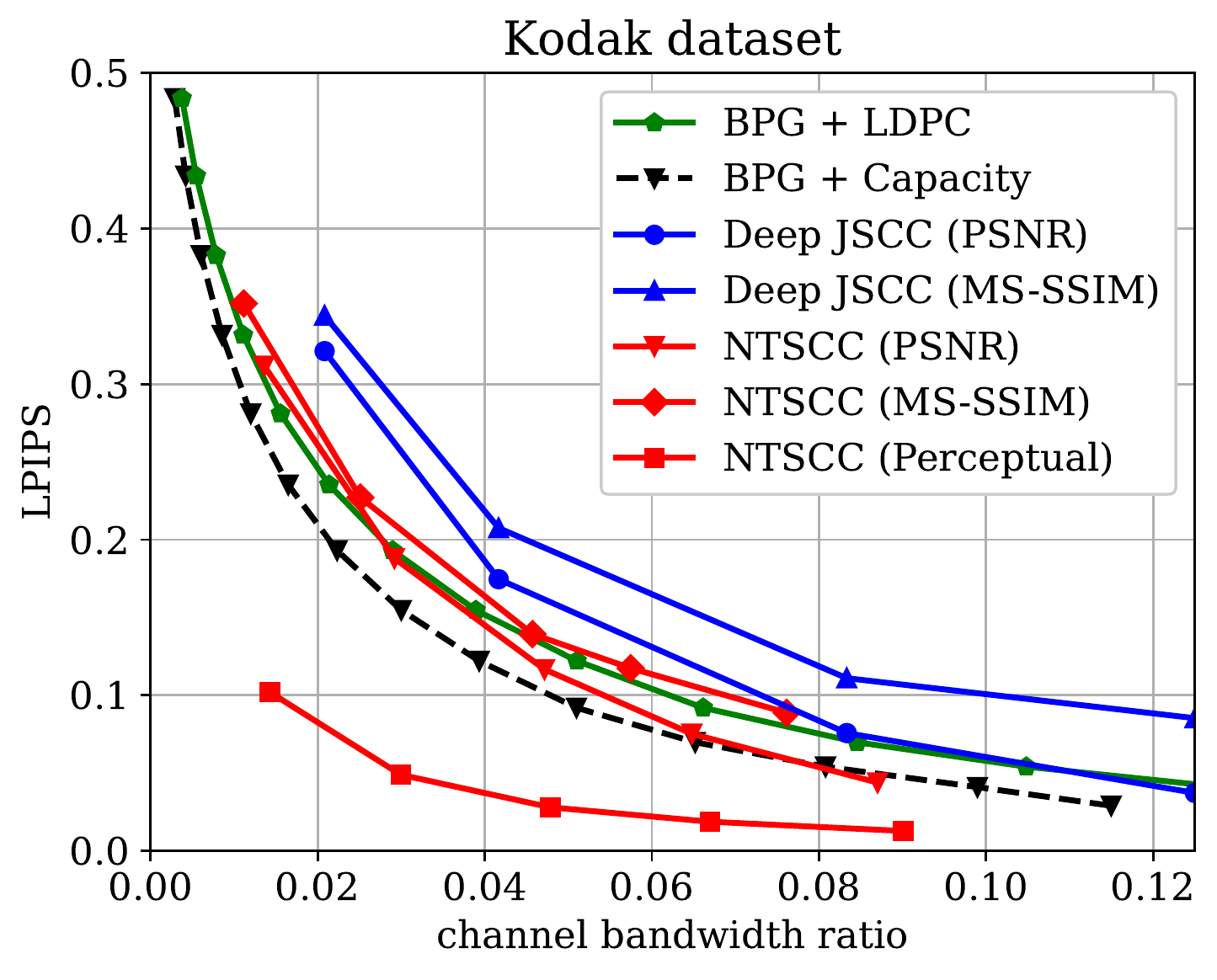}
		}
		\hspace{.50in}
		\quad
		\subfigure[]{
			\includegraphics[scale=0.47]{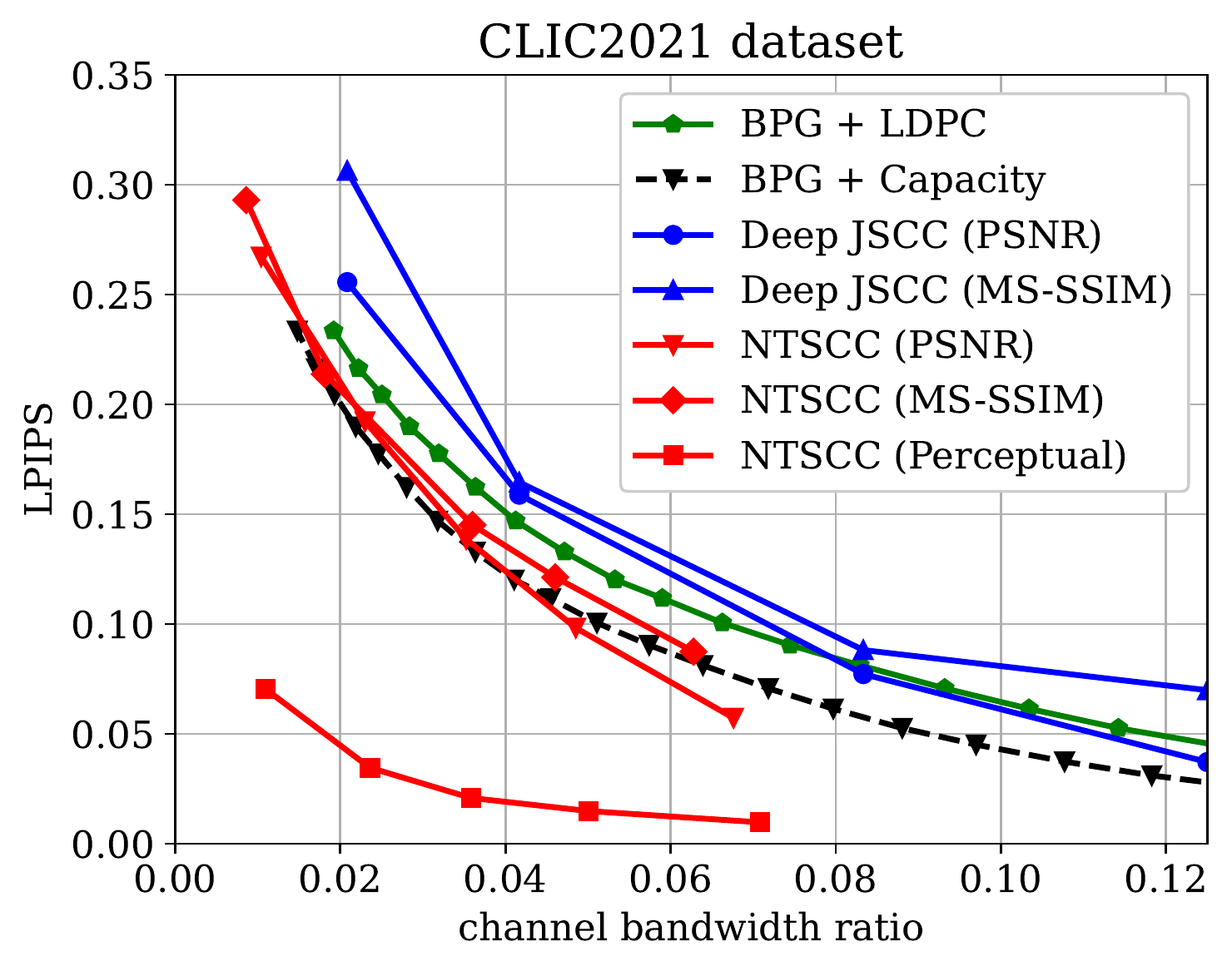}
		}
		\caption{LPIPS perceptual performance versus the average CBR over the AWGN channel at $\text{SNR} = 10$dB.}
		\label{Fig12}
	\end{center}
\end{figure*}

\makeatletter
\renewcommand{\@thesubfigure}{\hskip\subfiglabelskip}
\makeatother

\begin{figure*}
\vspace{-1em}
	\begin{subtable}
		\centering
		\small
		\begin{tabular}{m{0.174\textwidth}m{0.1307\textwidth}<{\centering}m{0.1307\textwidth}<{\centering}m{0.1307\textwidth}<{\centering}m{0.1307\textwidth}<{\centering}m{0.1307\textwidth}<{\centering}}
			& Original & BPG + LDPC &  Deep JSCC (\textsc{PSNR}) &  NTSCC (PSNR) &  NTSCC (Perceptual)  \\
		\end{tabular}
	\end{subtable}

	\begin{center}

   \hspace{-.05in}
	\subfigure[$512\times768$] {\includegraphics[width=0.15\textwidth]{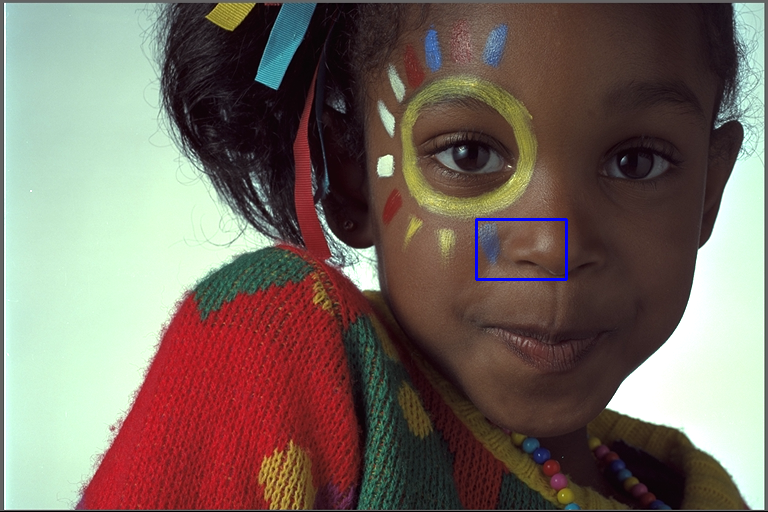}}
	\hspace{-.10in}
	\quad
	\subfigure[$R$ / PSNR (dB)] {\includegraphics[width=0.15\textwidth]{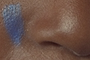}}
	\hspace{-.20in}
	\quad
	\subfigure[0.043 (\textit{0\%}) / 34.22]{\includegraphics[width=0.15\textwidth]{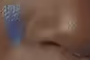}}
	\hspace{-.20in}
	\quad
	\subfigure[0.042 (\textcolor{blue}{\textit{--2\%}}) / 31.10] {\includegraphics[width=0.15\textwidth]{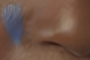}}
	\hspace{-.20in}
	\quad
	\subfigure[0.038 (\textcolor{blue}{\textit{--12\%}}) / 34.22] {\includegraphics[width=0.15\textwidth]{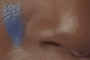}}
	\hspace{-.20in}
	\quad
	\subfigure[0.039 (\textcolor{blue}{\textit{--9\%}}) / 33.62]{\includegraphics[width=0.15\textwidth]{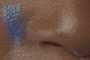}}

    \hspace{-.05in}
	\subfigure[$512\times768$] {\includegraphics[width=0.15\textwidth]{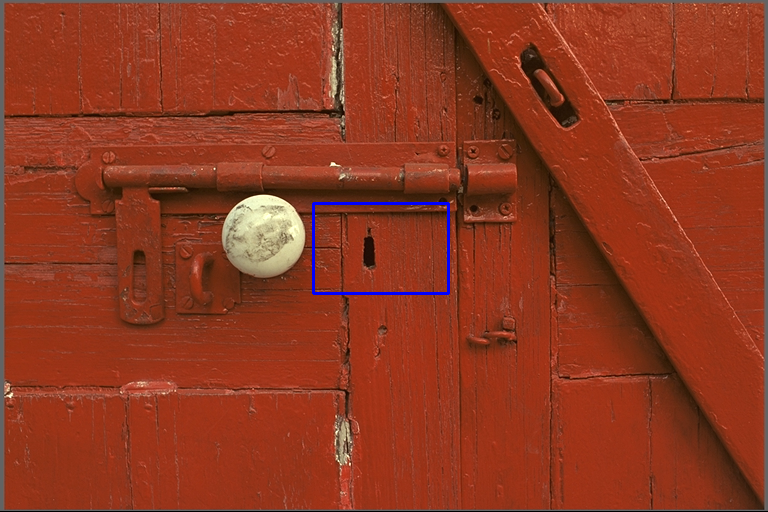}}
	\hspace{-.10in}
	\quad
	\subfigure[$R$ / PSNR (dB)] {\includegraphics[width=0.15\textwidth]{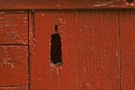}}
	\hspace{-.20in}
	\quad
	\subfigure[0.023 (\textit{0\%}) / 31.97]{\includegraphics[width=0.15\textwidth]{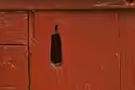}}
	\hspace{-.20in}
	\quad
	\subfigure[0.042 (\textcolor{red}{\textit{+83\%}}) / 31.71] {\includegraphics[width=0.15\textwidth]{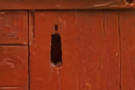}}
	\hspace{-.20in}
	\quad
	\subfigure[0.022 (\textcolor{blue}{\textit{--4\%}}) / 32.27] {\includegraphics[width=0.15\textwidth]{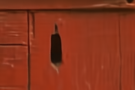}}
	\hspace{-.20in}
	\quad
	\subfigure[0.023 (\textit{0\%}) / 31.46]{\includegraphics[width=0.15\textwidth]{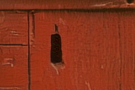}}
	
	\hspace{-.05in}
	\subfigure[$512\times768$] {\includegraphics[width=0.15\textwidth]{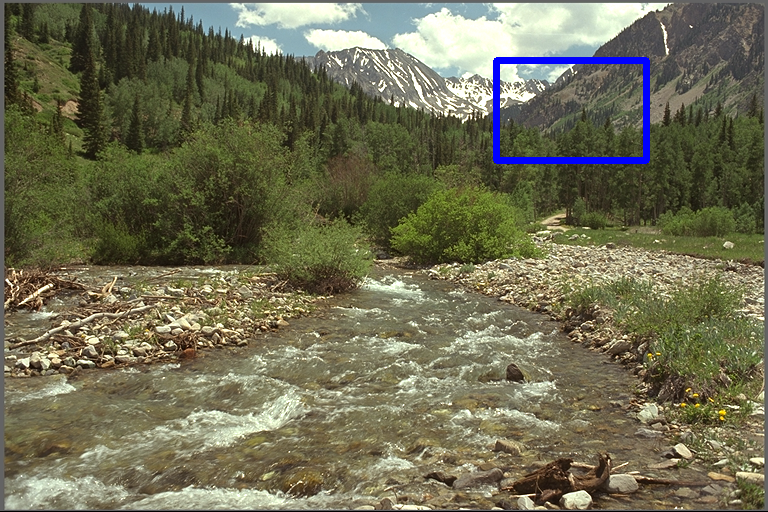}}
	\hspace{-.10in}
	\quad
	\subfigure[$R$ / PSNR (dB)] {\includegraphics[width=0.15\textwidth]{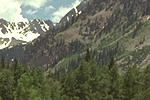}}
	\hspace{-.20in}
	\quad
	\subfigure[0.029  (\textit{0\%}) / 23.12]{\includegraphics[width=0.15\textwidth]{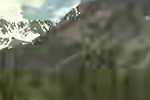}}
	\hspace{-.20in}
	\quad
	\subfigure[0.042 (\textcolor{red}{\textit{+45\%}})  / 23.67] {\includegraphics[width=0.15\textwidth]{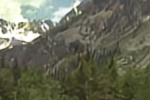}}
	\hspace{-.20in}
	\quad
	\subfigure[0.026 (\textcolor{blue}{\textit{--10\%}}) / 23.44] {\includegraphics[width=0.15\textwidth]{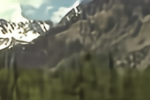}}
	\hspace{-.20in}
	\quad
	\subfigure[0.027 (\textcolor{blue}{\textit{--7\%}}) / 22.99]{\includegraphics[width=0.15\textwidth]{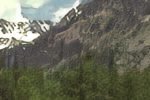}}

	\hspace{-.05in}
	\subfigure[$1512\times2016$] {\includegraphics[width=0.15\textwidth]{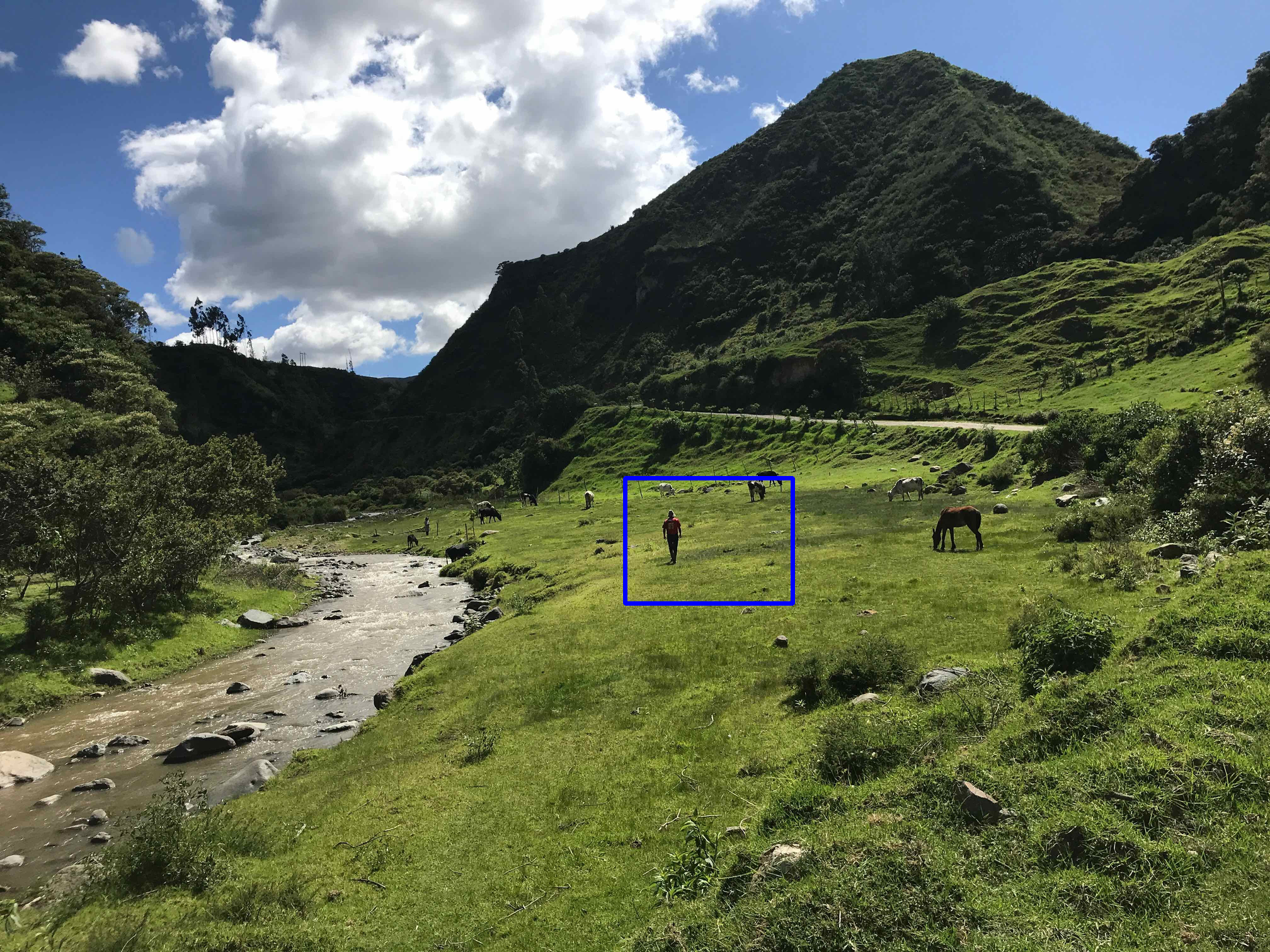}}
	\hspace{-.10in}
	\quad
	\subfigure[$R$ / PSNR (dB)] {\includegraphics[width=0.15\textwidth]{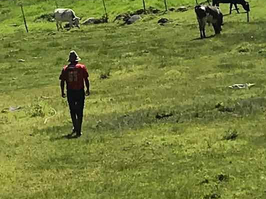}}
	\hspace{-.20in}
	\quad
	\subfigure[0.020  (\textit{0\%}) / 23.71]{\includegraphics[width=0.15\textwidth]{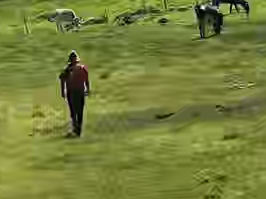}}
	\hspace{-.20in}
	\quad
	\subfigure[0.021 (\textcolor{red}{\textit{+5\%}})  / 25.09] {\includegraphics[width=0.15\textwidth]{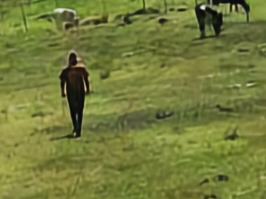}}
	\hspace{-.20in}
	\quad
	\subfigure[0.020 ({\textit{0\%}}) / 26.67] {\includegraphics[width=0.15\textwidth]{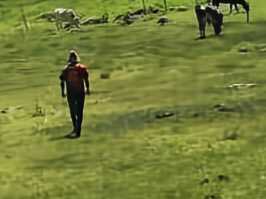}}
	\hspace{-.20in}
	\quad
	\subfigure[0.018 (\textcolor{blue}{\textit{--10\%}}) / 25.50]{\includegraphics[width=0.15\textwidth]{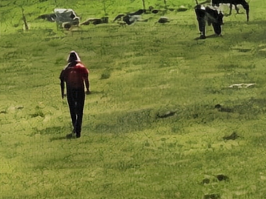}}

	\caption{Examples of visual comparison. The first column shows the original image. The second column shows the cropped patch in original frame. The third to the sixth column show the reconstructed images by using different transmission schemes over the AWGN channel at $\text{SNR} = 10$dB, where the metrics in parentheses indicate the model training target loss function (PSNR and Perceptual). Red number and blue number indicate the percentage of bandwidth cost increase and saving compared to the baseline ``BPG + LDPC'' scheme.}
\label{Fig13}
	\end{center}
\vspace{-2em}
\end{figure*}

\begin{figure*}

	\begin{subtable}
	\centering
	\small
	\begin{tabular}{m{0.174\textwidth}m{0.1307\textwidth}<{\centering}m{0.1307\textwidth}<{\centering}m{0.1307\textwidth}<{\centering}m{0.1307\textwidth}<{\centering}m{0.1307\textwidth}<{\centering}}
		& Original & BPG + LDPC &  Deep JSCC (MS-SSIM) &  NTSCC (MS-SSIM) &  NTSCC (Perceptual)  \\
	\end{tabular}
	\end{subtable}
	\begin{center}

	\hspace{-.05in}
	\subfigure[$1499\times1002$] {\includegraphics[width=0.15\textwidth]{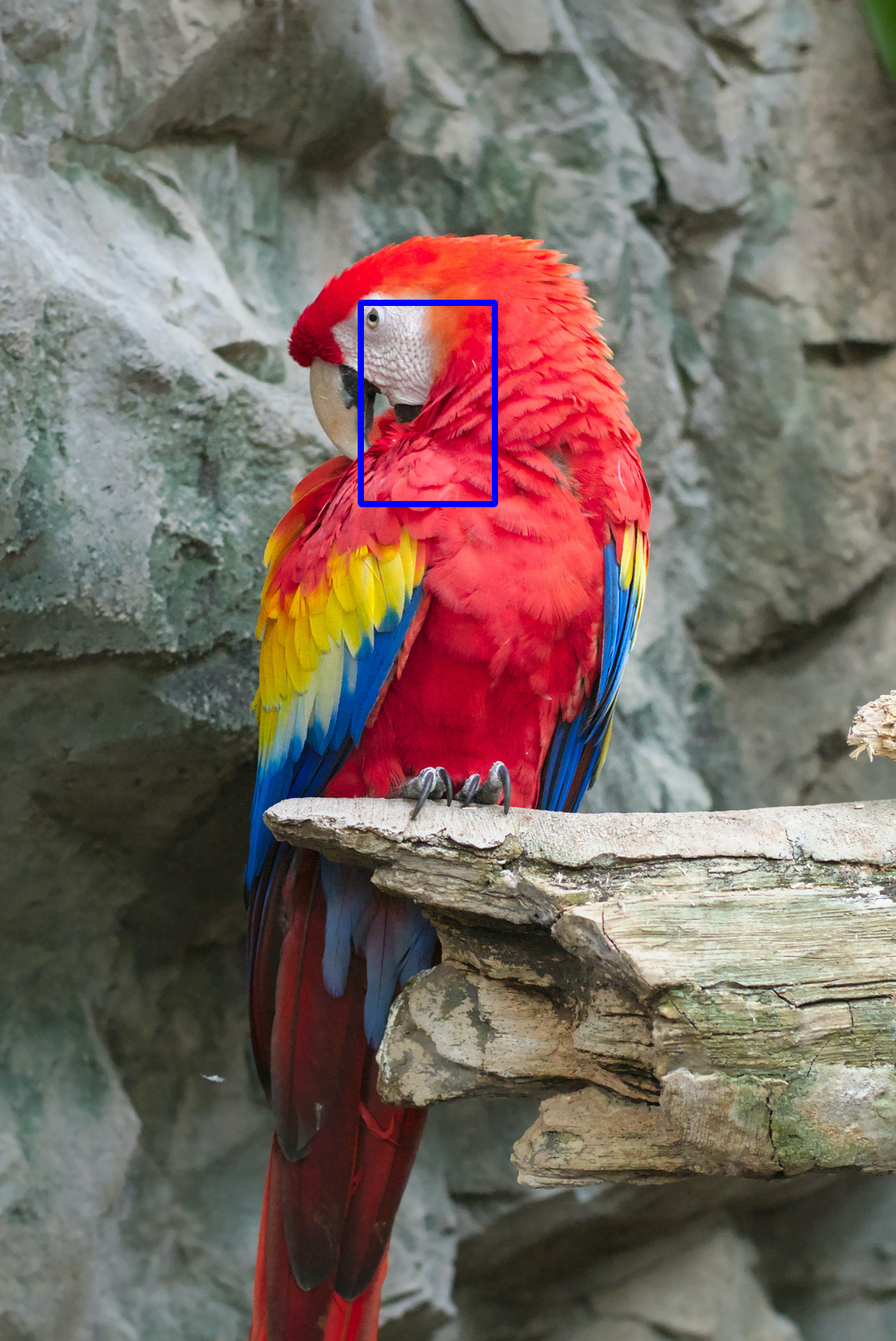}}
	\hspace{-.10in}
	\quad
	\subfigure[$R$ / MS-SSIM (dB)] {\includegraphics[width=0.15\textwidth]{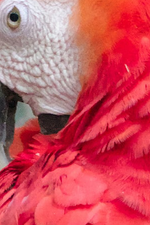}}
	\hspace{-.20in}
	\quad
	\subfigure[0.022 (\textit{0\%}) / 12.84]{\includegraphics[width=0.15\textwidth]{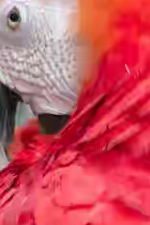}}
	\hspace{-.20in}
	\quad
	\subfigure[0.021 (\textcolor{blue}{\textit{--5\%}}) / 11.67] {\includegraphics[width=0.15\textwidth]{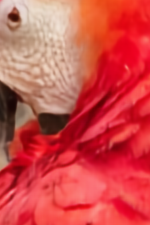}}
	\hspace{-.20in}
	\quad
	\subfigure[0.019 (\textcolor{blue}{\textit{--14\%}}) / 14.78] {\includegraphics[width=0.15\textwidth]{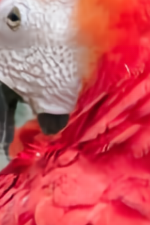}}
	\hspace{-.20in}
	\quad
	\subfigure[0.022 (\textit{0\%}) / 15.45]{\includegraphics[width=0.15\textwidth]{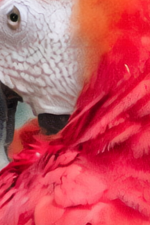}}
	
	\hspace{-.05in}
	\subfigure[$2048\times1365$] {\includegraphics[width=0.15\textwidth]{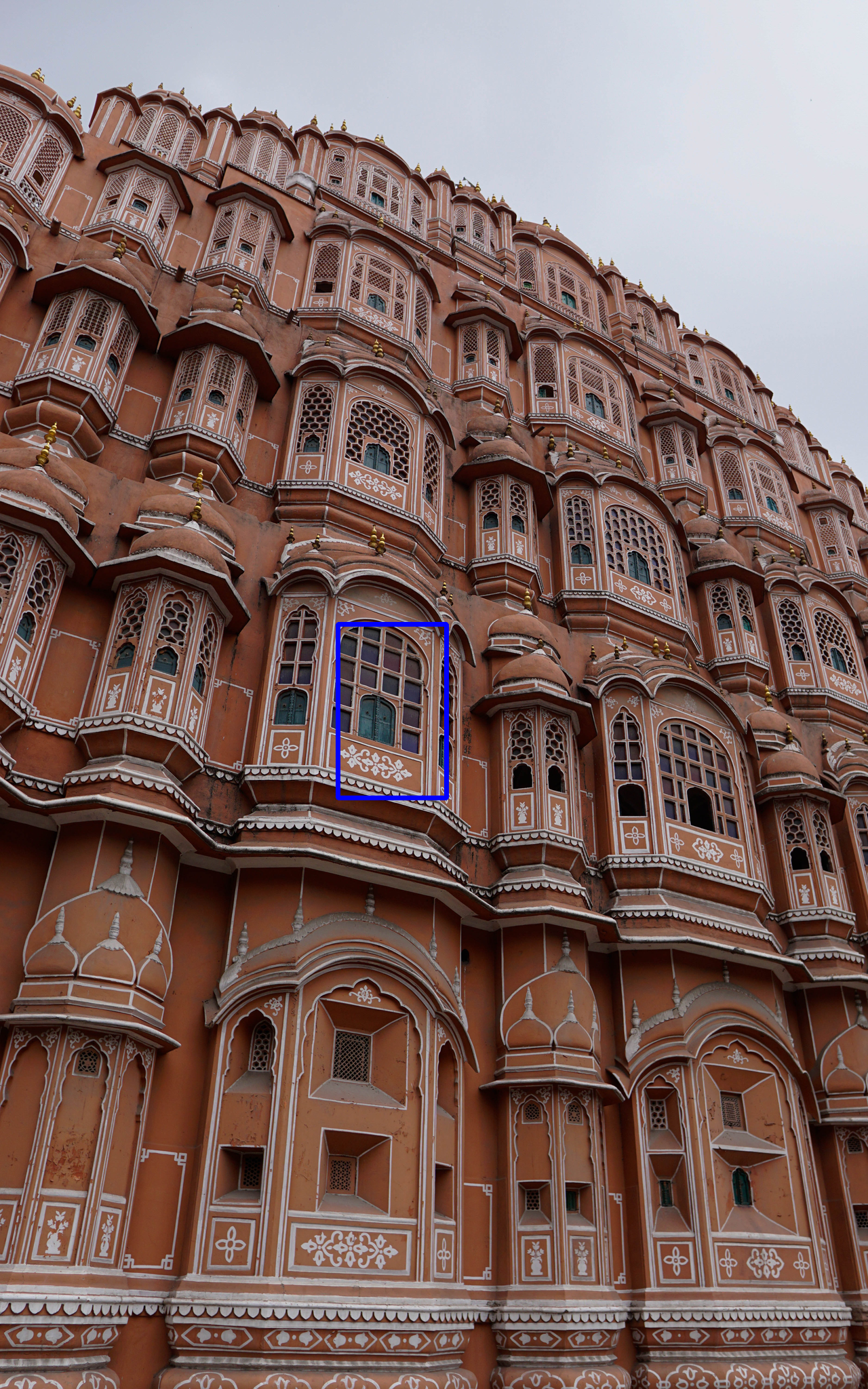}}
	\hspace{-.10in}
	\quad
	\subfigure[$R$ / MS-SSIM (dB)] {\includegraphics[width=0.15\textwidth]{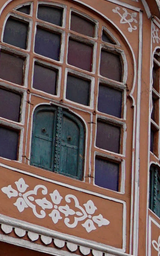}}
	\hspace{-.20in}
	\quad
	\subfigure[0.042 (\textit{0\%}) / 14.93]{\includegraphics[width=0.15\textwidth]{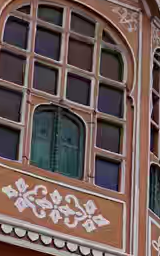}}
	\hspace{-.20in}
	\quad
	\subfigure[0.042 (\textit{0\%}) / 15.85] {\includegraphics[width=0.15\textwidth]{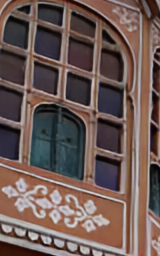}}
	\hspace{-.20in}
	\quad
	\subfigure[0.025 (\textcolor{blue}{\textit{--40\%}}) / 14.89] {\includegraphics[width=0.15\textwidth]{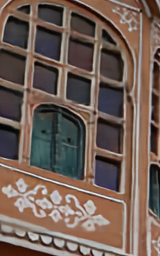}}
	\hspace{-.20in}
	\quad
	\subfigure[0.033 (\textcolor{blue}{\textit{--21\%}}) / 16.68]{\includegraphics[width=0.15\textwidth]{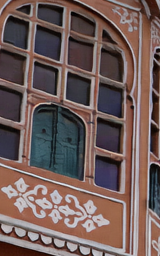}}
	
    \caption{Examples of visual comparison, where the configurations are the same as that in Fig. \ref{Fig13} except for alternating PSNR as MS-SSIM.}
\label{Fig14}
	\end{center}
\vspace{-1em}
\end{figure*}

Fig. \ref{Fig11}(a) and Fig. \ref{Fig11}(b) show the RD performance in terms of MS-SSIM at the AWGN channel with $\text{channel SNR} = 10$dB. Since MS-SSIM yields values between 0 (worst) and 1 (best), and most values are higher than 0.9, we converted the MS-SSIM values in dB to improve the legibility. We focus on the high-resolution images. The NTSCC models are trained with $\lambda \in \{1, 1/4, 1/16, 1/32, 1/128\}$ and $\eta = {0.4}$, a lower value of $\lambda$ leads to a larger bandwidth ratio $R$. The difference in the value of hyperparameters $\lambda$ and $\eta$ stems from the difference in the loss function value. Results indicate that the proposed NTSCC method outperforms ``BPG + Capacity'' and deep JSCC schemes by a large margin, and it achieves a greater improvement on high-resolution images and high CBR regions. Compared to the PSNR results, we can find that the BPG series are inferior to the learning-based schemes because BPG compression is designed to be optimized for squared error with hand-selected constraints \cite{HIFIC}.

We also provide BD-CBR and BD-MS-SSIM results in Table \ref{Table2}. It can be observed that in the high-resolution CLIC2021 dataset, the proposed NTSCC can save up to 64.27\% CBR, while deep JSCC only saves 31.77\%. Similar improvements can be found in Fig. \ref{Fig11}(c) and Fig. \ref{Fig11}(d), which demonstrate the MS-SSIM values with a range of channel SNRs.

\subsubsection{LPIPS Performance}

Apart from the above PSNR and MS-SSIM distortion metrics, regarding the goal of semantic communications, we further use the human-perception orientated LPIPS loss function \cite{lpips} to train the NTSCC model. This learning-based metric is aligned with the perceptual quality. In addition, as aforementioned, when the NTSCC model targets at human perceptual metric, the NTSCC distortion term in the loss function for model training refers to \eqref{eq_lpips_loss}. Fig. \ref{Fig12} shows the LPIPS results versus the average CBR, a lower LPIPS indicates a lower distortion. For learning-based schemes (deep JSCC and NTSCC), we mark PSNR and MS-SSIM in parentheses to indicate the model training target. The ``NTSCC (Perceptual)'' curves in Fig. \ref{Fig12} indicate the performance of NTSCC model optimized for learned perceptual distortion in \eqref{eq_lpips_loss}. Clearly, the perceptually optimized NTSCC overpasses other schemes by a large margin.

To intuitively demonstrate the effect of perceptual optimization, we further pick visible results on the testset as shown in Fig. \ref{Fig13} and Fig. \ref{Fig14}. From these examples, we can observe the proposed NTSCC model learned under perceptual loss can achieve higher visual quality with much lower channel bandwidth cost. In particular, it avoids artifacts effectively and produces a high-fidelity reconstruction with more generated details, while the traditional ``BPG + LDPC'' scheme exhibits blocking artifacts. Therefore, the proposed NTSCC method can better support future semantic communications.

\subsection{Ablation Study}

For the ANN architectures of NTSCC, we have proposed using the Transformers to replace the CNN backbone. In addition, we have designed the rate adaption mechanism based on dynamic networks. Besides, the receiver further utilizes hyper-prior to refine the latent codes. To verify the effectiveness of these methods, we report the transmission performance in the following settings.

\begin{enumerate}[(1)]
 \item We take off the refinement operation in $f_d^{\star}$ by directly feeding the tentatively recovered version $\mathbf{\check{y}}$ into the synthesis transform module. In this case, we remove the corresponding cost of side information $\mathbf{z}$ since it will not be used in the receiver. This ablation study aims to verify whether the hyperprior $\mathbf{\bar z}$ needs to be transmitted as the side information.

 \item We further invalidate the rate adaption module in $f_e$ and $f_d^{\star}$ by using a unified channel bandwidth cost for each $y_i$ and deleting the additional rate tokens. Since the network cannot learn a trade-off between channel bandwidth ratio and distortion, it is trained end-to-end with the MSE loss. In this case, the network indeed falls back to a deep JSCC network with a vision Transformer (ViT) backbone.
\end{enumerate}

\begin{figure}[t]
	\setlength{\abovecaptionskip}{0.cm}
	\setlength{\belowcaptionskip}{-0.cm}
	\centering{\includegraphics[scale=0.47]{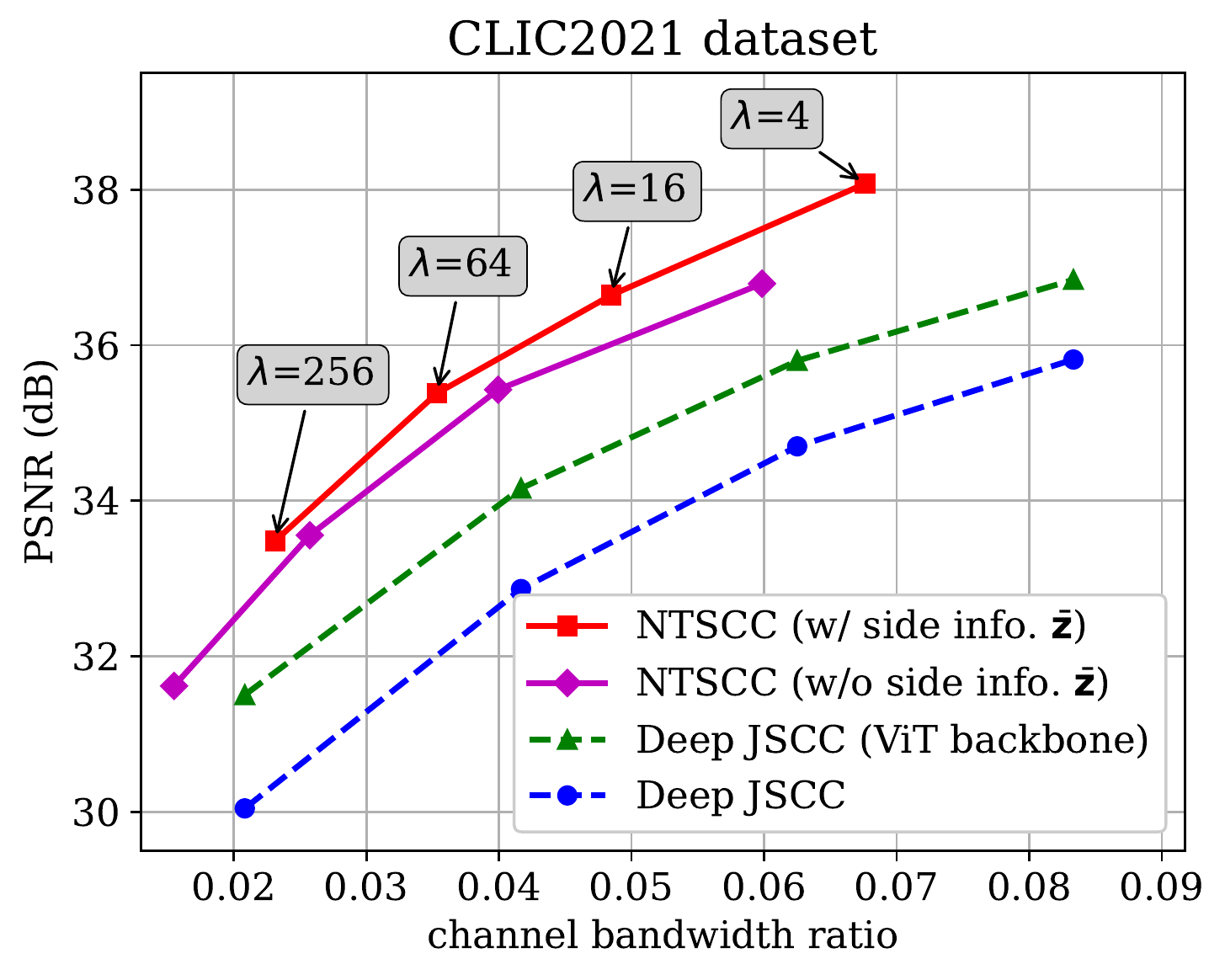}}
	\caption{Ablation study results in terms of the PSNR performance over the AWGN channel at $\text{SNR} = 10$dB, and $\eta = 0.2$ for the NTSCC model.}\label{Fig15}
	\vspace{0em}
\end{figure}

\begin{table}[t]
\renewcommand{\arraystretch}{1.3}
  \centering
  \small

  \caption{Amount of side information as a function of total CBR.}

  \begin{tabular}{!{\vrule width1pt}m{0.4cm}|m{1.4cm}|m{2.6cm}|m{2.6cm}!{\vrule width1pt}}

    \Xhline{1pt}

     \centering $\lambda$ & \centering Total CBR & \centering CBR of side info. $\mathbf{\bar z}$ & \centering CBR of side info. $\mathbf{\bar k}$ \tabularnewline

     \Xhline{1pt}

     \centering $256$ & \centering $0.023$ & \centering $0.0057$ ($24.65\%$) & \centering $0.0015$ ($6.51\%$) \tabularnewline

     \hline

     \centering $64$ & \centering $0.035$ & \centering $0.0068$ ($19.30\%$) & \centering $0.0015$ ($4.26\%$) \tabularnewline

     \hline

     \centering $16$ & \centering $0.048$ & \centering $0.0072$ ($14.75\%$) & \centering $0.0015$ ($3.11\%$) \tabularnewline

     \hline

     \centering $4$ & \centering $0.068$ & \centering $0.0090$ ($13.28\%$) & \centering $0.0015$ ($2.23\%$) \tabularnewline

     \Xhline{1pt}

  \end{tabular}
  \label{Table3}
\end{table}

Fig. \ref{Fig15} shows the transmission PSNR performance under these settings. Compared to the benchmark deep JSCC with CNN backbone, the performance gain of ``Deep JSCC (ViT backbone)'' verifies the benefit of exploiting vision Transformer network architecture as a stronger backbone. The scheme of ``NTSCC without (w/o) side info. $\mathbf{\bar z}$'' further shows a clear performance gain versus the ``Deep JSCC (ViT backbone)'', which verifies the coding gain brought by the proposed rate adaptive transmission mechanism.

Comparing the intact ``NTSCC with (w/) side info. $\mathbf{\bar z}$'' with the scheme of ``NTSCC w/o side info. $\mathbf{\bar z}$'', the performance gain verifies the effect of the proposed hyperprior-aided codec refinement mechanism. The CBR of each scheme counts up the total bandwidth costs to transmit codewords $\mathbf{s}$ and all types of side information, thus, all comparisons are fair. For the intact scheme of ``NTSCC w/ side info. $\mathbf{\bar z}$'', Table \ref{Table3} shows the amount of side information as a function of total CBR. The CBR of hyperprior side information $\mathbf{\bar z}$ grows with the total CBR. The CBR of side information $\mathbf{\bar k}$ indicating the allocated channel bandwidth of each patch embedding $y_i$ stays identical, and it is quite lower than the total CBR. With respect to the results of Fig. \ref{Fig15} and Table \ref{Table3}, we conclude that exploiting the hyperprior side information $\mathbf{\bar z}$ at the receiver can indeed obtain performance gain while the bandwidth cost for transmitting $\mathbf{\bar z}$ is relatively high. We can also adopt the NTSCC architecture without transmitting the hyperprior side information $\mathbf{\bar z}$, due to the lack of decoder refinement on $f_d$, some performance degradation can be observed, but its performance still greatly surpasses traditional deep JSCC schemes. Therefore, the use of hyperprior side information transmission link in Fig. \ref{Fig4} is optional for practical implementations of NTSCC.

\section{Conclusion}\label{section_conclusion}

This paper has proposed a new class of high-efficiency joint source-channel coding methods that can learn to closely adapt to the source distribution under the nonlinear transform. It was collected under the name ``NTSCC''. Unlike traditional deep JSCC methods, NTSCC first learns a nonlinear analysis transform to map the source data into the latent space, then transmits the latent representation to the receiver via a group of learned variable-length neural JSCC encoders and a noisy channel. In this way, although the source distribution cannot be known by the transceiver in advance, the proposed NTSCC model can learn the source latent representation, as well as an entropy model on the latent representation to implicitly approximate the true source distribution. Accordingly, novel adaptive rate transmission and hyperprior-aided codec refinement mechanisms have been developed to improve end-to-end transmission performance effectively. For the NTSCC model optimization, both the widely-used PSNR/MS-SSIM distortion metrics and the emerging human perceptual metric LPIPS have been adopted, matching the goal of end-to-end semantic communications. In summary, this paper has proposed a promising method to enable the elaborate design of learning-based joint source-channel coding.

\bibliographystyle{IEEEbib}
\bibliography{myRef}

\end{document}